\documentclass[pre,twocolumn,twoside,byrevtex,superscriptaddress,floatfix]{revtex4-1}
\usepackage[export]{adjustbox}
\usepackage{settings}
\usepackage[table,pdftex,dvipsnames]{xcolor}    % loads also »colortbl«
\setboolean{twocolswitch}{true}
\newcommand{\eps}{\varepsilon}
\newcommand{\dist}[1]{\mathbf{#1}}
\newcommand{\usageRate}{f}
\newcommand{\intUsage}{I}

\definecolor{todoblue}{RGB}{0, 91, 187}

\begin{document}
  \rowcolors{2}{gray!25}{white}
\title{\protect  Hurricanes and hashtags: \\
 Characterizing online collective attention for natural disasters}
\author{
\firstname{Michael V.}
\surname{Arnold}
}
\email{michael.arnold@uvm.edu}

\affiliation{MassMutual Center of Excellence in Complex Systems and Data Science,
Computational Story Lab, Vermont Complex Systems Center, and
the Department of Mathematics and Statistics,
University of Vermont,
Burlington, Vermont 05405, USA}
%%%%%%%%%%%%%%%%%%%%%%%%%%%%%%%%%%%%%%%%%%%%%

%%%%%%%%%%%%%%%%%%%%%%%%%%%%%%%%%%%%%%%%%%%%%
\author{
\firstname{David Rushing}
\surname{Dewhurst}
}
%\email{David.Dewhurst@uvm.edu}

\affiliation{MassMutual Center of Excellence in Complex Systems and Data Science,
Computational Story Lab, Vermont Complex Systems Center, and
the Department of Mathematics and Statistics,
University of Vermont,
Burlington, Vermont 05405, USA}
\affiliation{
MassMutual Data Science,
Boston, Massachusetts 02110, USA
}
%%%%%%%%%%%%%%%%%%%%%%%%%%%%%%%%%%%%%%%%%%%%%

%%%%%%%%%%%%%%%%%%%%%%%%%%%%%%%%%%%%%%%%%%%%%
\author{
\firstname{Thayer}
\surname{Alshaabi}
}

\affiliation{MassMutual Center of Excellence in Complex Systems and Data Science,
Computational Story Lab, Vermont Complex Systems Center, and
the Department of Mathematics and Statistics,
University of Vermont,
Burlington, Vermont 05405, USA}
%%%%%%%%%%%%%%%%%%%%%%%%%%%%%%%%%%%%%%%%%%%%%

%%%%%%%%%%%%%%%%%%%%%%%%%%%%%%%%%%%%%%%%%%%%%
\author{
\firstname{Joshua R.}
\surname{Minot}
}

\affiliation{MassMutual Center of Excellence in Complex Systems and Data Science,
Computational Story Lab, Vermont Complex Systems Center, and
the Department of Mathematics and Statistics,
University of Vermont,
Burlington, Vermont 05405, USA}
%%%%%%%%%%%%%%%%%%%%%%%%%%%%%%%%%%%%%%%%%%%%%

%%%%%%%%%%%%%%%%%%%%%%%%%%%%%%%%%%%%%%%%%%%%%
\author{
\firstname{Jane L.}
\surname{Adams}
}

\affiliation{MassMutual Center of Excellence in Complex Systems and Data Science,
Computational Story Lab, Vermont Complex Systems Center, and
the Department of Mathematics and Statistics,
University of Vermont,
Burlington, Vermont 05405, USA}
%%%%%%%%%%%%%%%%%%%%%%%%%%%%%%%%%%%%%%%%%%%%%

%%%%%%%%%%%%%%%%%%%%%%%%%%%%%%%%%%%%%%%%%%%%%
\author{
\firstname{Christopher M.}
\surname{Danforth}
}
%\email{cdanfort@uvm.edu}

\affiliation{MassMutual Center of Excellence in Complex Systems and Data Science,
Computational Story Lab, Vermont Complex Systems Center, and
the Department of Mathematics and Statistics,
University of Vermont,
Burlington, Vermont 05405, USA}
%%%%%%%%%%%%%%%%%%%%%%%%%%%%%%%%%%%%%%%%%%%%%

%%%%%%%%%%%%%%%%%%%%%%%%%%%%%%%%%%%%%%%%%%%%%
\author{
\firstname{Peter Sheridan}
\surname{Dodds}
}
%\email{pdodds@uvm.edu}
\affiliation{MassMutual Center of Excellence in Complex Systems and Data Science,
Computational Story Lab, Vermont Complex Systems Center, and
the Department of Mathematics and Statistics,
University of Vermont,
Burlington, Vermont 05405, USA}

\date{\today}

\begin{abstract}
  \protect
  We study collective attention paid towards hurricanes through the lens of $n$-grams on Twitter, a social media platform with global reach.
Using hurricane name mentions as a proxy for awareness, we find that the exogenous temporal dynamics are remarkably similar across storms, but that overall collective attention varies widely even among storms causing comparable deaths and damage. 
We construct `hurricane attention maps' and observe that hurricanes causing deaths on (or economic damage to) the continental United States generate substantially more attention in English language tweets than those that do not.  
We find that a hurricane’s Saffir-Simpson wind scale category assignment is strongly associated with the amount of 
attention it receives.
Higher category storms receive higher proportional increases of attention per proportional increases in number of deaths or dollars of damage, than lower category storms.
The most damaging and deadly storms of the 2010s, Hurricanes Harvey and Maria, generated the most attention and were remembered the longest, respectively.
On average, a category 5 storm receives 4.6 times more attention than a category 1 storm causing the same number of deaths and economic damage.
 
\end{abstract}

\pacs{89.65.-s,89.75.Da,89.75.Fb,89.75.-k}

%% 89.65.-s	Social and economic systems
%% 89.75.Da	Systems obeying scaling laws
%% 89.75.Fb	Structures and organization in complex systems
%% 89.75.-k	Complex systems (for complex chemical systems, see 82.40.Qt; for biological complexity, see 87.18.-h)

\maketitle

%%%%%%%%% end of author(s), address(es) plus abstract

%% add sections here...
%%%%%%%%%%%%%%%%%%%%%%%%%%%%%%%%%%%%%%%%%%%%%%%%%%%%%%%%%%%
\section{Introduction}\label{sec:introduction}
The collective understanding and memory of historic events shapes the common world views of societies.
In a narrative economy, attention is a finite resource generating intense competition~\cite{Shiller:2017iu, shiller2019narrative, Leskovec:2009cfa, tufekci2013not, franck2019economy,humphreys2009construction, citton2017ecology,Franck1999Science,nowak2017social}. 
As commerce and communication shift to online platforms, so too has the narrative economy moved to the digital realm. 
In 2018, over \$100 billion dollars were spent on internet advertising in the United States, nearly overtaking the \$110 billion spent on traditional media advertising---about 1\% of the US GDP~\cite{IAB:2018}.
Today, social media both facilitates and records an extraordinary percentage of the world's public communication~\cite{newman2009rise,perrin2015social}. 
For computational social scientists, the migration of parts of the narrative economy to the web continues to present an immense opportunity, as the discipline becomes data-rich~\cite{Michel:2011gma,pechenick2015a}. 
%% Previous work in the field of `culturomics' such as Google's $n$-grams create and %% quantify large scale text corpora from books~\cite{Michel:2011gma}, but until %% %% ecently, recording or measuring the words of any but a tiny minority of influential %% individuals was technologically infeasible.

Academics have become interested in narrative spreading around newsworthy events on social media platforms such as Twitter,
as increasingly political fights for influence or narrative control
are fought by actors as wide ranging from activists and police departments~\cite{Gallagher:2018gi},
to state censors suppressing discourse internally
and  state supported troll factories
spreading divisive narratives internationally~\cite{cit:qYG945fL, Colleoni2014, Gruzd2014, Barbera2015, Broniatowski2018, Subrahmanian2016}. In 2019, the social media platform Twitter boasted over 145 million daily active users~\cite{Salinas2019Oct}.

%%%%%%%%%%%%%%%%%%%%%%%%%%%
% subsubsection: attention on Twitter
Quantifying the spread of narratives and the total attention commanded by them is a daunting task.
Recent work has made progress in tracking the spread of quoted and modified phrases through the news cycle, 
and others have worked to identify actant-relationships and compile contextual story graphs from social media posts~\cite{Leskovec:2009cfa,Shahbazi:2019ub}.
In comparison, quantifying attention directed towards a topic, person or event is a somewhat easier task.
Rather than identifying actors and identifying what they act on, as is the case for narrative attention, we can simply count mentions of an entity. 
Since increasing raw attention or number of mentions is often the zeroth order activity in public relations campaigns, 
quantifying the volume of attention, irrespective of the sentiment or narrative within which the attention is embedded, seems a natural first step~\cite{Dodds:2019to}.

% users of the idiom 'much ink has been spilled...' will have to adapt to this being a falsifiable claim.
%%%%%%%%%%%%%%%%%%%%%%%%%%%
% subsubsection: previous work on quantifying the dynamics of attention.
An understanding of attention has typically focused on time dynamics as measured by the number of mentions in a given corpus,
explaining either temporal decay of interest or heavy-tailed allocation of attention given to a spectrum of topics through some preferential attachment mechanism.
 ~\cite{Dorogovtsev:2000jd,Golosovsky:2012ho,Valverde:2007fs,Higham:2017ja,Higham:2017dz,Wang:2013bq,Candia:2019gd,LorenzSpreen:2019im}. 
 Another group of studies have worked to classify attention time series from social media as either exogenous or endogenous to the system, 
 modeling the functional form of collective attention decay,
or determining if spreading crosses a critical threshold~\cite{Crane:2008hm,Lehmann:2012fj,Wu:2007eja,dilankile:1tENHRzB}. 
While these studies have typically focused on scientific works, patents, or cultural products such as movies,
the rise of large social media datasets have enabled the investigation of a wider range of topics in online public discourse~\cite{Ladle:2016ka}.

In this study we examine the collective attention focused on hurricanes, using Twitter, which allows us to capture more natural speech intended for human readers as opposed to search terms.
Twitter data has been used to measure shifts in collective attention surrounding exogenous events like earthquakes by looking for jumps in the Jensen-Shannon divergence between tweet rate distributions between days, or creating real-time earthquake detection using keyword based methods~\cite{Sasahara:2013e,Sakaki2010}.

Here, we use collective attention in a more narrow sense. Instead of looking for anomalous tweet rates, we study $n$-gram usage rates for hashtags and $2$-grams associated with individual events. 
Specifically, we examine the usage rates of hashtags and $2$-grams matching the case-insensitive pattern ``\texttt{\#hurricane*}'' and ``\texttt{hurricane *}'', respectively. 
Natural disasters provide an ideal case study, since they are generally unexpected, producing the signature of an exogenous event.
However, the volume of attention given to any particular hurricane varies widely across several orders of magnitude, 
as does the severity of the storm in terms of the lives lost and damages caused.

 Prior efforts have examined the attention received by disasters by type and location, as measured by time devoted on American television news network coverage, and striking discrepencies: for example, to have the same estimated probability of news coverage as a disaster in Europe, a disaster in Africa would need to cause 45 times as many deaths~\cite{Eisensee:gu}. The same study found that in order to receive equivalent coverage to a deadly volcano, a flood would need to cause 674 times as many deaths, a drought 2,395 times as many, and a famine 38,920 times as many casualties.

Strong hurricanes are more likely to capture attention than weak hurricanes, and hurricanes impacting the continental United States capture much more attention than those failing to make landfall. To what degree does attention shrink when hurricanes make landfall outside of the continental US?
%\todo{Please frame sentences that state beliefs as ``we believe that'' since this is encouraged in the Bayesian framework in which you analyze data here.}
The 2017 hurricane season is a particularly stark example, showing that for comparably powerful storms above category 4, those projected to make landfall over the continental United States were talked about nearly an order of magnitude more than Hurricane Maria, which impacted Puerto Rico, and two orders of magnitude more than Hurricane Jose, which never made landfall.

%\todo{DRD: Recommend not providing a summary of findings in the intro. That's what the abstract is for. Instead, just say what you are going to do in the rest of the paper. That will make the intro much shorter.}
%Generally storms causing more deaths and damage are talked about more, and storms with slow or inefficient relief efforts are talked about longer.
%Further investigation into the scaling relationship between hurricane impact measures and the attention received by the storms, showed that not only are more powerful storms talked about more than less powerful storms, but higher category storms receive a larger proportional increase in attention per proportional increase in impact inflicted than lower category storms: A 10-fold increase in the number of deaths caused by a category 5 hurricane is associated with a 26-fold increase in hashtag usage frequency, while for a category 1 or category 2 storm is associated with 2- and 2.5-fold increases in attention respectively. 

Given the attention received by some hurricanes so unbalanced, we must ask the question: Do government or humanitarian relief resources get dispersed with greater generosity for storms that capture public attention, or are these organizations insulated from popular attention? For the 2017 hurricane season, more money was spent more quickly to aid the victims of hurricanes Harvey and Irma than victims of Hurricane Maria, contributing to the significantly higher death toll and adverse public health outcomes in Puerto Rico~\cite{Willison:2019gz}. While the attention and policies of government agencies are not usually dictated from Twitter, public attention certainly has some effect on the focus of agencies and allocation of government resources, and recently more attention has been focused on understanding the discourse on social media before, during, and after natural disasters~\cite{allen2018president,Niles2019,Cody2017, ahmed2020construction}

We structure our paper as follows. In \cref{sec:methods}, we outline our methods and data sources, covering the collection of $n$-gram usage rate data in English tweets as well as data sources for hurricane locations and impacts. 
In \cref{sec:results}, we examine the spatial associations between hurricanes and the attention they receive, we compute and compare measures of total attention, maximum daily attention, and non-parametic measures of the rate of attention decay for the most damaging hurricanes in the past decade. We present conclusions in  \cref{sec:discussion}.  

\section{Methods}\label{sec:methods}

\subsection{n-gram usage rates}
We query the daily usage rate of hashtags referencing hurricanes are queried from a corpus of $1$-gram---words or other single word-like constructs---usage rate time series, computed from approximately 10\% of all posts (``tweets'') from 2009 to 2019 collected from Twitter's ``decahose''
\cite{limuch}.
We define usage rate, $\usageRate$, as
$$\usageRate(t) = c_\tau(t)\bigg/\sum_{\tau' \in \mathcal{D}_t}c_{\tau'}(t),$$
with count, $c_\tau$, of a particular $1$-gram divided is by the count of all $1$-grams occurring on a given day, $\mathcal{D}_t$.
The usage rates are 
based only on the usage rate of $1$-grams observed in tweets classified as English by FastText, a language classification tool~\cite{Joulin:2016uo, Alshaabi2020a}. %\todo{update citation for Thayer's paper}
Our usage rate data set includes separate usage rates for $1$-grams in ``organic'' tweets, 
tweets that are originally authored,
as well as usage rates of $1$-grams in all tweets (including retweets and quote tweets). More details about the parsing of the Twitter $n$-gram data set are available in
\cite{Alshaabi2020b}.

For the purpose of studying attention, our usage rates are derived from the corpus with all tweets,
including retweeted text, to better reflect not only the number of people tagging a storm, but also the number of people who decide the information contained therein was worth sharing.

We studied the usage rate of $1$-grams exactly matching the form ``\texttt{\#hurricane*}'',
where \texttt{*} represents a storm's name. 
We also measured the usage rate of 
$2$-grams matching the pattern ``\texttt{hurricane *}'' for each storm name.  All string matching is case-insensitive.

For the ten years covered by the HURDAT2 dataset overlapping with our Twitter dataset, there have been 75 storms reaching at least category $1$ in the North Altlantic Basin. Within our 10\% sample of tweets, we count over all storms a total of 1,824,842 hashtag usages within a year of each storm, and 3,643,411 instances of the matching $2$-gram.

\subsection{Deaths, damages, and locations}
To augment our usage rate data set, we downloaded data associated with all hurricanes in the North Atlantic basin from 2008 to 2019 from Wikipedia~\cite{wiki:xxx}.
Included in the Wikipedia data are the damage estimates (US\$) and deaths caused by each storm, as well as the dates of activity and areas effected. We also used the HURDAT2 data set containing the positions and various meteorological attributes of all North Atlantic hurricanes from 1900 to 2018 for the spatial component of this work~\cite{Weinkle:wr}. 
For the time range overlapping with the Twitter derived data set, HURDAT2 has 3 hour resolution.

\section{Results and Discussion}\label{sec:results}

\subsection{Hurricane Attention Maps}
\begin{figure*}[ht]
  \centering	
    \includegraphics[width=0.95\linewidth]{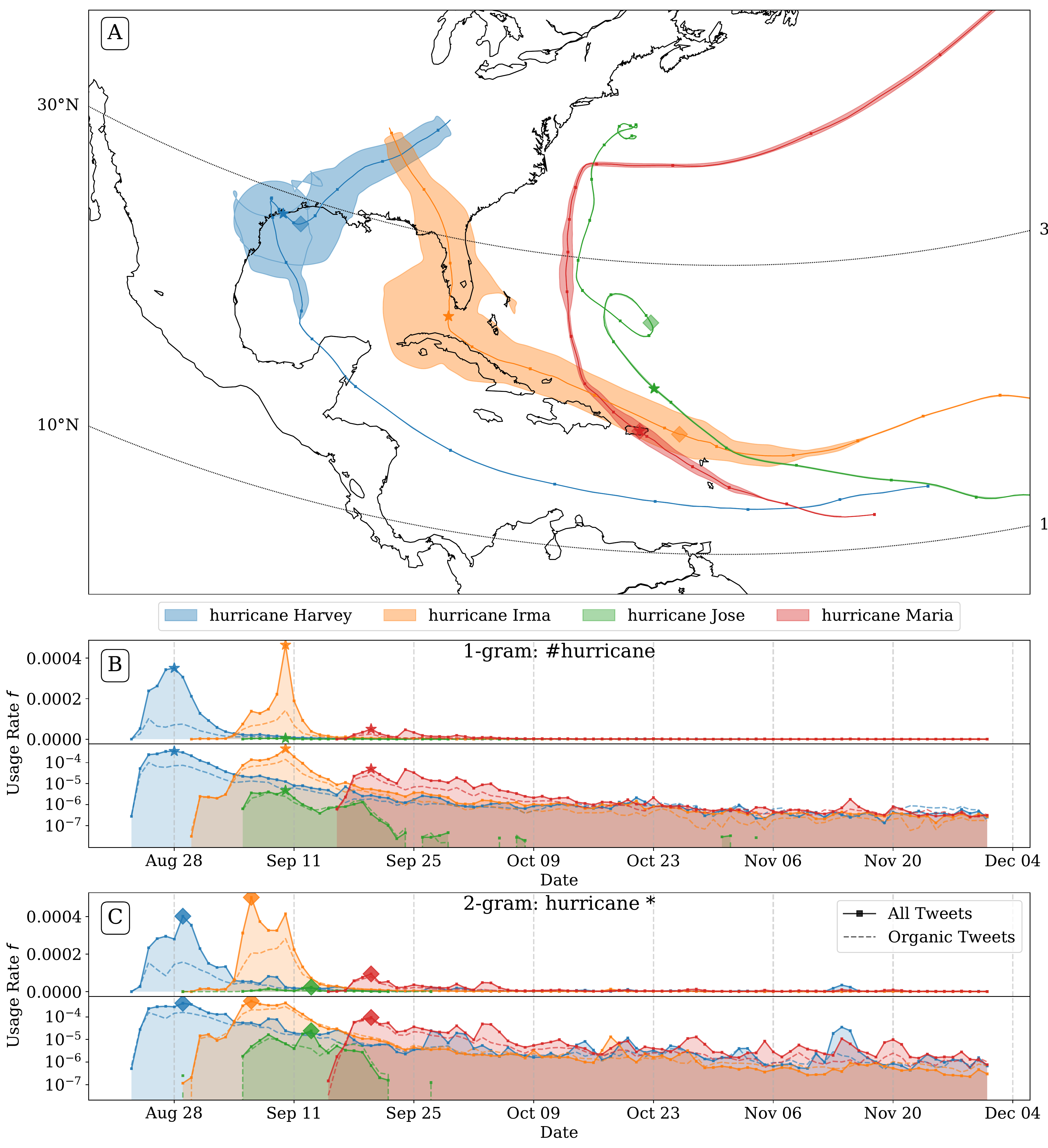}
  \caption{\textbf{ Hashtag attention map and usage rate time series}
    for $1$-grams matching the case-insensitive pattern ``\texttt{\#hurricane*}''  for all four hurricanes reaching at least category 4 in the 2017 hurricane season.
    Markers along the hurricane trajectory indicate the National Oceanic and Atmospheric Administration (NOAA) reported position for every day at noon UTC. 
    On the map, the smoothed rate of hashtag usage is wrapped in an envelope around the hurricane trajectory in panel A, showing the spatial dependence of attention on Twitter. In the lower two plots, panels B and C, we show the usage rates for hashtags and $2$-grams matching \texttt{hurricane*} in English language tweets on linear and logarithmic scales. 
    Usage rates within all tweets are indicated with a solid line, while usage rates in `organic' tweets (tweets that are not retweets), are represented by a  dashed line. The day of maximum attention on Twitter is marked with a star or a diamond for hashtags or 2-grams, respectively.
    Generally, hurricanes making landfall on the continental United States received greater attention than those not making landfall. The hashtag usage rate for urricanes Harvey and Irma at their maximum were approximately an order of magnitude larger than the maximum hashtag usage corresponding to hurricane Maria, and two orders of magnitude larger than Hurricane Jose.}\label{fig:tracks}
\end{figure*}
%\todo{try grid based attention maps with krieging. Can do spanish and english comparison}

In \cref{fig:tracks}, we show hurricane positions as well as their hashtag usage rate timeseries with a time series indicating the usage rate of the hashtag of the form \texttt{\#hurricane*}.

We plot the same hashtag usage rate time series below on both linear and logarithmic axes, as well as $2$-gram usage rates. For clarity, we only include hurricanes reaching at least category 4.

The hurricane map tracks are meant to show the spatial dependence of attention given to hurricanes, while giving enough visual cues to connect locations along the path to the time the attention was observed.  
We generated the map shown in \cref{fig:tracks} by filling in the polygon defined by
the set of points lying at the end of a line segment of length proportional to the smoothed usage rate of the related hashtag,
along the vector normal to the current velocity of the hurricane, and centered at the hurricane position at the given time.

Our hashtag usage rate is at the day scale, while HURDAT has 3 hour resolution,
so the wrapped attention volume is smoothed with a moving average with a window size of one day to avoid discontinuous jumps.
This method obscures any sub-day scale resolution on the map, which could be related to the daily fluctuation of tweet volume as well as varying interest in the hurricanes. While we lose some granularity using daily usage rates, the decays in attention are spread out over days and weeks for smaller storms, and months for larger storms. Daily resolution is sufficient to capture the longer decays in attention, which are our primary interest.

Examining the map, we can see the minimal attention paid to Hurricane Harvey as it traveled across the Caribbean sea and made landfall in Mexico. It is only after crossing the Gulf of Mexico that the hashtag registered on our instrument, and only when it was about to make landfall over Texas did the hashtag usage rate approach its maximum rate, approximately 3 of every 10,000 $1$-grams in English tweets. 
It appears that the devastation wrought by Harvey primed hurricane-related conversation, as the next hurricane, Irma was talked about long before it made landfall. 
While Irma was talked about with a similar usage rate as Harvey as it impacted Puerto Rico, Hispaniola, and Cuba, it spiked while making landfall in the Florida keys.

Comparing the attention generated by the previous two storms, Hurricane Maria generated substantially less hashtag usage. 
The peak of its attention gathered as it made landfall over Puerto Rico as a category 4 storm, with less than a fifth of the attention as the hurricanes making landfall on the US.
Part of the reason may be due the affected area being Spanish speaking, while our hashtag usage measurement only counts occurrences in English tweets. 
We find that usage rates of the $2$-gram ``\texttt{Huracán Maria}'' in Spanish tweets were also lower than the usage rates for  ``\texttt{Huracán Irma}'', but comparable to those for ``\texttt{Huracán Harvey}.'' See \cref{fig:2017_attention_share_es} to compare top hurricane related $2$-gram time series for the 2017 hurricane season in English and Spanish.

Another potential contributing factor for the low volume of Hurricane Maria tweets could be that Puerto Rico's electric grid was destroyed and 95\% of cell towers were down in the aftermath of the storm, making it impossible for those directly affected to communicate about the storm \cite{Scott2018}. 
%\todo{Citations please!}
Unfortunately, due to Twitter's usage norms in this time period, we do not have locations for the vast majority of tweets.
The number of people affected by the storms could also help explain the different levels of attention, as both Hurricane Harvey and Irma affected 19 million people, while Maria affected about 4 million \cite{Hurricanes2019}.

%It would be interesting to see the geographic distribution of users using the hashtags associated with the hurricanes. If the majority of the people tweeting about the storm are in its path or at least in the cone of possible paths, it might explain the disparity of attention generated by each. If the majority of people tweeting about the storm are not in its path, then perhaps the hashtag frequency is reflective of the importance of the story to the users. Obviously, the severity of the storm itself is not the determining factor, since hurricanes that stayed as see, but were equally as powerful were essentially ignored, while those effecting lives and livelihoods received considerable attention.    

\subsection{Hurricane Attention Comparison}
%%% FIGURE 2 %%%
\begin{figure*}[ht!]
  \centering	
    \includegraphics[width=1\linewidth]{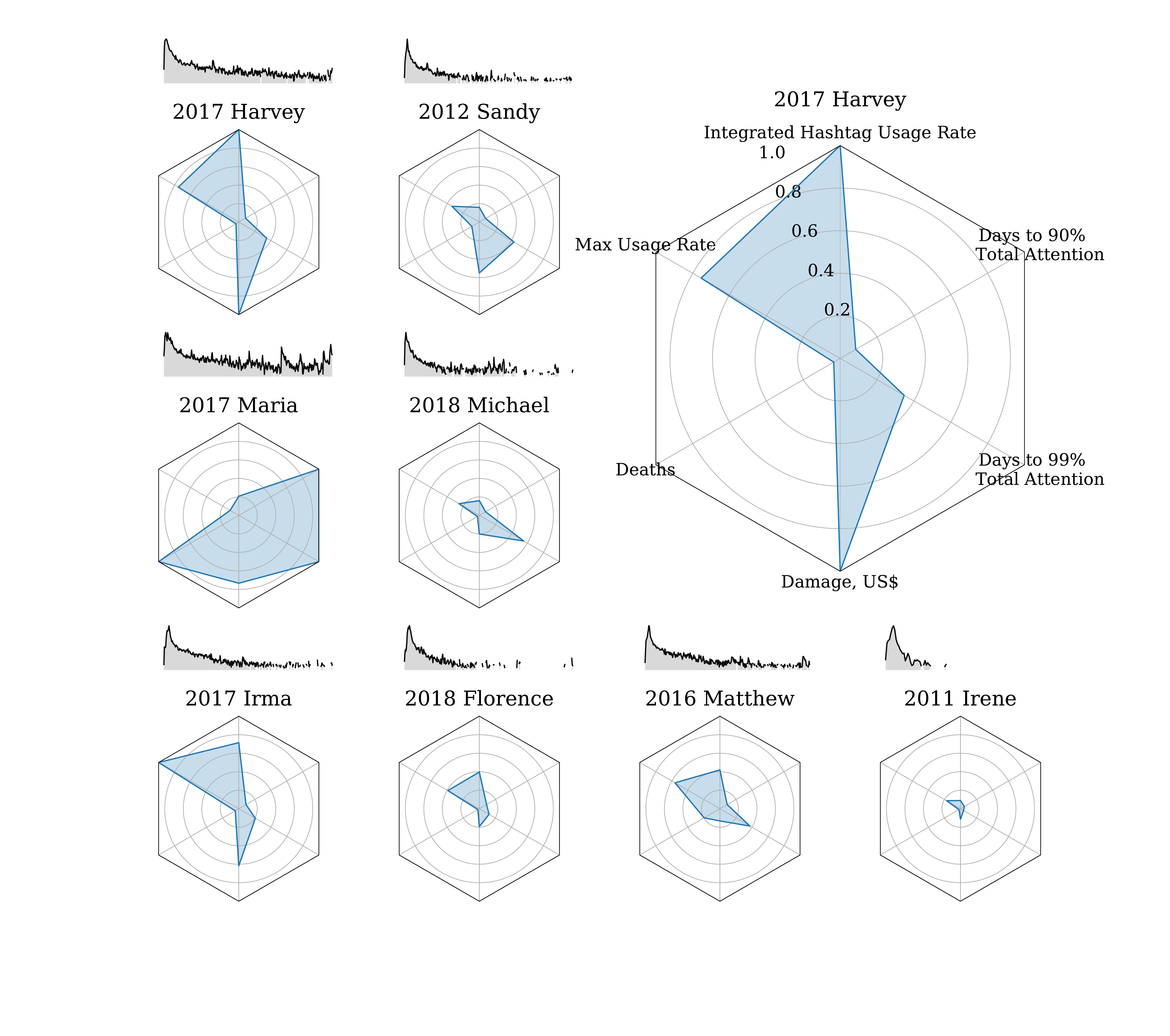}  
  \caption{
    \textbf{Radar plots comparing the eight most monetarily damaging hurricanes in the North Atlantic basin from 2009 to 2018.} For each plot, starting at the top position and rotating clockwise the measures are: the sum of usage rate of the hashtag, the number of days to reach 90\% and 50\% of the total attention received during that season, the total cost in dollars attributed to damage caused by the hurricane (in its year), the number of deaths attributed to the hurricane, and maximum  usage rate of the hashtag during the year of interest. All measurements are normalized to the maximum value achieved by any hurricane. Hurricane Harvey was the most talked about hurricane, as well as the most damaging. Hurricane Irma was the most talked about on any single day. 
    Hurricane Maria caused the most deaths, and had the longest attention half-life of all measured hurricanes.
    Raw values for this figure are shown in \cref{tab:hurricane_compare}. Hashtag usage rate spark lines above each radar plot are normalized to show the common decay shape, and can not be compared to evaluate \textit{relative} volume, and are shown on a log scale.
  }
  %half-lives fit from a biexponential model of the decay in frequency from the max frequency $y = c_1e^{-\lambda_1 t}+ c_2 e^{-\lambda_2 t}$
  \label{fig:radar}
\end{figure*}  

To compare the variation in attention received by different storms, we combined measurements of the hashtag usage rate with deaths and damages caused by each storm from 2009 to 2019. 
The supplimentary materials, \cref{tab:hurricane_compare}, shows these raw measured values for the most damaging hurricanes in this period.

In \cref{fig:radar}, we show radar plots (radial, categorical charts) comparing six measurements of impact and attention for each of the eight most damaging hurricanes in the time period of study \cite{Liu2008}. 
Included measurements are:
\begin{itemize} 
\item Max Usage Rate---peak attention on any single day
\item Integrated Usage Rate---total attention over the entire hurricane season
\item Quantile 0.9: $Q_{0.9}$---days to 90\% attention
\item Quantile 0.99: $Q_{0.99}$---days to 99\% attention
\item Damage---total damage caused by the storm in US dollars 
\item Deaths---total deaths associated with the storm (both direct and indirect)
\end{itemize}
The relative magnitude of each quantity is shown as a fraction of the maximum value for any storm in the study. The quantile values are non-parametric measurements of the attention time scale---comparable to half-lives but without the assumption of an exponential decay.
Some storms receive significant interest months after they pass, usually related to the recovery efforts.  Spark lines above each plot show the attention time series for the year after each storm, as measured by the log usage rate, but do not convey relative scale. 

 %%%%%%%%%%%%%%% %%%%%%%%%%%%%%% begin new radar discussion section %%%%%%%%%%%%%%% %%%%%%%%%%%%%%%

% damage
The three most damaging storms, Hurricanes Harvey, Maria, and Irma, all destroyed tens of billions of dollars of property. Storms in \cref{fig:radar} are ordered by damage, with the least damaging being Hurricane Irene in 2011, which still destroyed  an estimated \$14 billion in property.

The most deadly North Atlantic hurricane in the past decade was Hurricane Maria, killing over 3000 people over the course of the extended disaster. The next most deadly storms were Hurricanes Matthew, Sandy, Irma, and Harvey, all killing at least 100 people. Among the storms shown in the \cref{fig:radar}, Hurricanes Florence and Irene were the least deadly, causing 58 and 57 deaths, respectively.

%max attention
The highest hashtag usage rate on a single day was associated with Hurricane Irma, reaching $\max f_{\tau} = 4.6\times 10^{-4}$, or 4.6 of every 10,000 $1$-grams, as the storm made landfall over the Florida Keys. Other storms reached comparable single day usage rates, such as Hurricanes Harvey and Matthew, reaching $\max \usageRate = 3.5\times 10^{-4}$ and $\max \usageRate = 2.6\times 10^{-4}$, respectively. Within the top eight most damaging storms, the hashtag associated with Hurricane Maria had the lowest maximum usage rate. The hashtag ``\texttt{\#hurricanemaria}'' appeared only five times for every 100,000 $1$-grams as Maria made landfall in Puerto Rico.

%total attention
The highest integrated hashtag usage rate was associated with Hurricane Harvey, followed by Hurricanes Irma, Matthew, and Florence. The integrated hashtag usage rate for ``\texttt{\#hurricaneharvey}'', $\intUsage = 2.3 \times 10^{-3}$. Hashtags associated with Hurricanes Sandy and Irene had the total attention, with $\intUsage = 3.7 \times 10^{-4}$ and $\intUsage = 2.0 \times 10^{-4}$, respectively.

% quantiles
Due to the extended crisis in the aftermath of Hurricane Maria, the hashtag continued to be used at relatively high volumes even a year after the storm had passed, leading to much larger value for $Q_{0.9}$ of 175 days \cite{roman2019satellite,Zorrilla2017}.
Typical values for $Q_{0.9}$ were around 1--4 days, with more prolonged and damaging storms like Harvey in 2017 taking 15 days to reach 90\% total attention. In comparison no other storm took longer than 100 days to reach this benchmark. 
We chose the longer term attention timescale benchmark, $Q_{0.99}$, to describe how long until nearly all storm focused attention has passed.
We observe the hashtag associated with Hurricane Maria is the largest for this measurement as well, with $Q_{0.99}$ of 363 days, which should be interpreted as attention not dying away within a year, since we truncate the timeseries after one year.
Hurricane Michael, Sandy, and Harvey also have triple digit values for $Q_{0.99}$, as they continued to be talked about, albeit at much lower levels than their peak. Other storms quickly lose attention, such as Hurricane Irene, which took only 12 days to reach 99\% total attention.

% overall shapes
We observed variation in the overall radar plot shape. More recent storms have been more damaging and deadly, and we find higher measures of total attention and attention decay. A number of storms like Sandy, Michael, and Matthew have relatively higher values for both maximum usage rate and number of days to reach 99\% total attention. While there is significant variation in the magnitude of these measurements, the essential exogenous shape of the hashtag usage rate timeseries, $\usageRate$, is consistent.

\subsection{Attention and Impact Regressions by Category}
\label{subsec:category_models}
%%% FIGURE 3 %%%
\begin{figure*}[ht!]
  \centering	
    \includegraphics[width=0.99\linewidth]{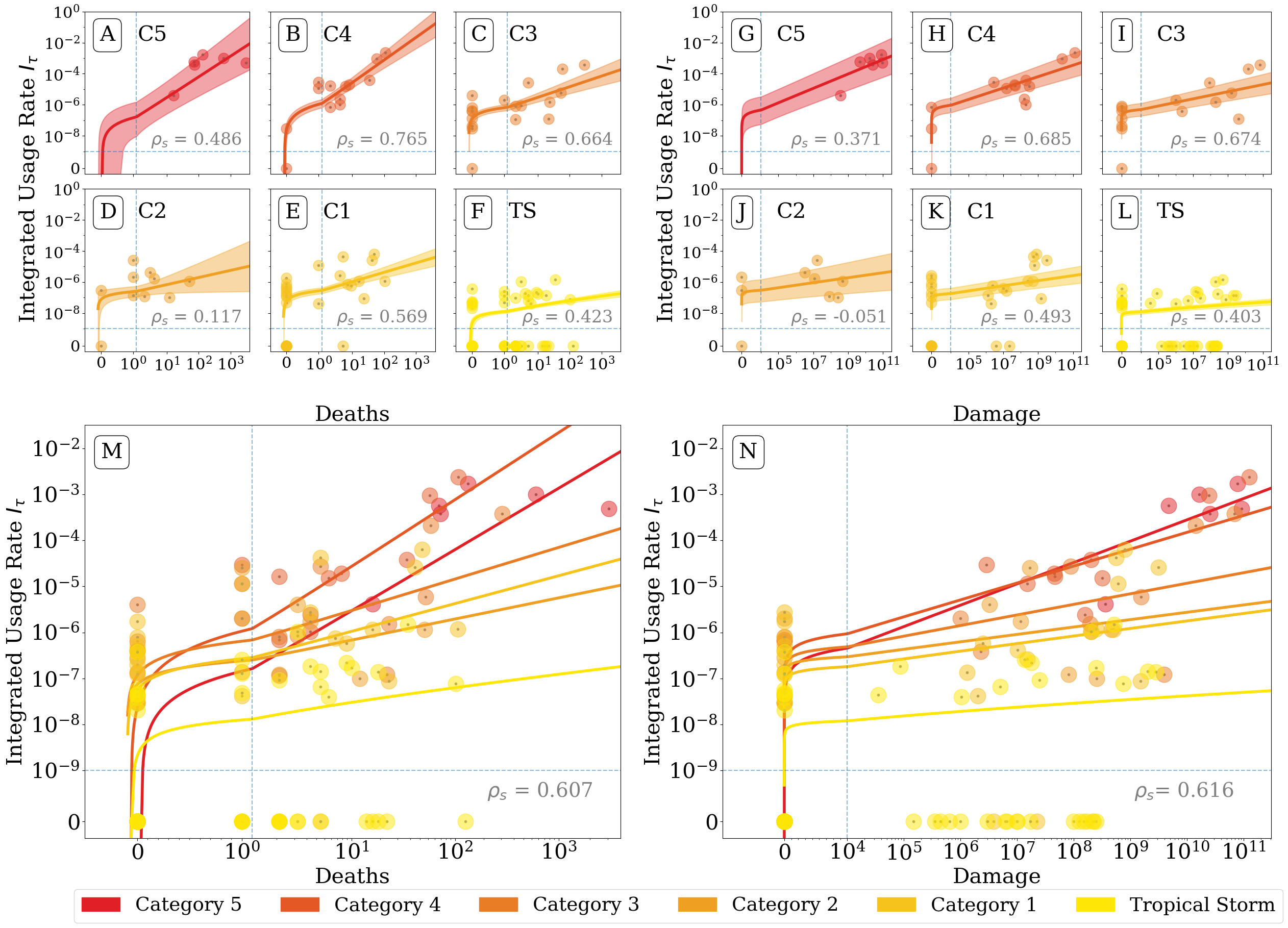}  
    
  \caption{
  \textbf{Scatter plots for integrated hashtag usage rate versus the deaths and damages caused by each storm.} There is a clear positive association between the total attention represented by hashtags and the impacts of these storms. We reported Spearman's rho, $\rho_s$, in the top left corner of each plot. While for some categories, there is little evidence for a positive association, for the entire dataset $\rho_s\sim 0.54$. We perform a Bayesian linear regression for each category storm between the  $\log \intUsage$ and log impacts. We show the mean model, along with the credible interval within a standard deviation of the mean model. We use hybrid axis with logarithmic scaling for most horizontal and vertical values and linear scaling near zero, in order to show storms that caused zero deaths or damages, as well as storms for which we measured a hashtag usage rate of zero. Changes in axis scaling occur at the blue dashed lines. Generally, more powerful storms received more attention, higher category storms received more attention even when causing minimal damage, and high category storms had a higher regression slope. These results suggest that for powerful storms, a given increase in impact was associated with a larger increase in attention. While for category one storms a 10-fold increase in deaths is associated with a two-fold increase in attention, for category five hurricanes, this same 10-fold increase in attention is associated with a 27-fold increase in attention.
  }
  \label{fig:bayesian regression}
\end{figure*}

We next explore the associations between damage, deaths, and attention given to hurricanes. 
In \cref{fig:bayesian regression}, we show the scaling relationship between attention and impacts for each category storm on the Saffir-Simpson wind scale \cite{taylor2010saffir}. 
Each sub-panel plots the integrated usage rate, $\intUsage = \sum_t \usageRate(t)$ for hashtag or $2$-gram $\tau$, 
against a measure of storm impact, where $t$ runs over an index of the 365 days after each storm began.
$\intUsage$ is chosen as a measure of total attention given to the storm during its respective hurricane season, which can be compared across years since it is already normalized to the total volume of conversation on Twitter.  
Color represents the maximum category storm reached, and the smaller subplots are breakout panels for each category. 
We include Spearman's $\rho$, a non-parametric measure of rank correlation, in each panel.

We perform linear regressions on storms in each category separately, a choice that models the attention received by different category storms as separate processes.
With models in \cref{subsec:combined_models}, we separately consider attention as a singular process where we account for the hurricane's maximum category rating using an explicit indicator variable.

\subsubsection{Model Choice and Fitting Procedure}

 For each category and each impact, we model total attention as 
\begin{equation}\label{eq:attention-linear-regression}
    \log_{10} \intUsage = a_0 +a_{\textnormal{impact}}X_{\textnormal{impact}} + \eps_{\tau},
\end{equation}
where $X_{\textnormal{impact}}$ is either $\log_{10}$ deaths or $\log_{10}$ damages caused by each storm.
We use a logarithmic model both to capture the scaling relationships between impacts and attention and to inform on the relative changes in attention associated with storm impacts.
We offset $\intUsage$ by $10^{-8}$ and the log impacts, $X_{\textnormal{impact}}$ by $\$10,000$ and $0.1$ deaths, respectively to avoid divergent log data where observed values are equal to zero.

We set a zero-centered normal prior on the slope of the regression model as $a_1 \sim \dist{normal}(0,1)$. 
We set a normal prior on the intercept of the model with mean equal to $\log_{10}\intUsage=-8$, the minimum value of the offset added to $\intUsage$.
We did not have strong beliefs about the likely precision of $a_0$ since it was not \textit{a priori} clear how much attention would be paid to hurricanes with very little associated monetary damage or few deaths.
We thus set a weak hyper-prior on the precision of $a_0$, $\tau\sim \dist{gamma}(3,1)$;
the intercept of the regression is distributed as 
$a_0 \sim \dist{normal}(-8, \tau^{-1})$.

We found regression coefficients by sampling with the No-U-Turn-Sampler (NUTS), using 8 chains with 2000 draws each after 1000 steps of burn-in \cite{hoffman2014no}. 
Our models converged, with the Gelman-Rubin statistic, $\hat{R}$, never exceeding 1.004 for any parameter in the 12 models fit.
%\todo{Please rewrite this with much more detail.  Also, how did the model converge? how do you know? i.e., cite Rhat statistics (Gelman-Rubin statistics) and 

\subsubsection{Model Posteriors and Discussion}

In \cref{fig:bayesian regression}, we show the fitted regressions for each category. 
The size of the impact and attention variables vary over many orders of magnitude,
but also include zero values, corresponding to storms that cause no deaths or damage,
or had zero usage of the hashtag associated with their name during the year the storm was active. 
Note that it should not be surprising that tropical storms appear to receive less attention via our hashtag usage rate measurement, since they never officially become hurricanes, and thus many of the tropical storm hashtags have an integrated usage rate, $\intUsage = 0$.

 To display all data, we use symmetric log axes: logarithmic for large values and linear for small values.
We indicate the switch point from linear to log space axis as blue dotted lines. 
This choice of axes causes the linear regressions on the log transformed data to appear curved for small values.

 In each of the small subplots of \cref{fig:bayesian regression}, we show the $1\sigma$ credible interval for the model as a band around the mean regression model.
The credible interval is noticeably wider for category five storms, which is reasonable given there are only seven storms reaching this category. 
Generally the mean regression lines are ordered such that higher category storms are receiving more attention than lower category storms.
The slopes of the regressions are also higher for higher category storms.
However, to better understand the models, we need to compare the model parameters individually.

\begin{figure*}[ht!]
  \centering	
\includegraphics[width=0.9\linewidth]{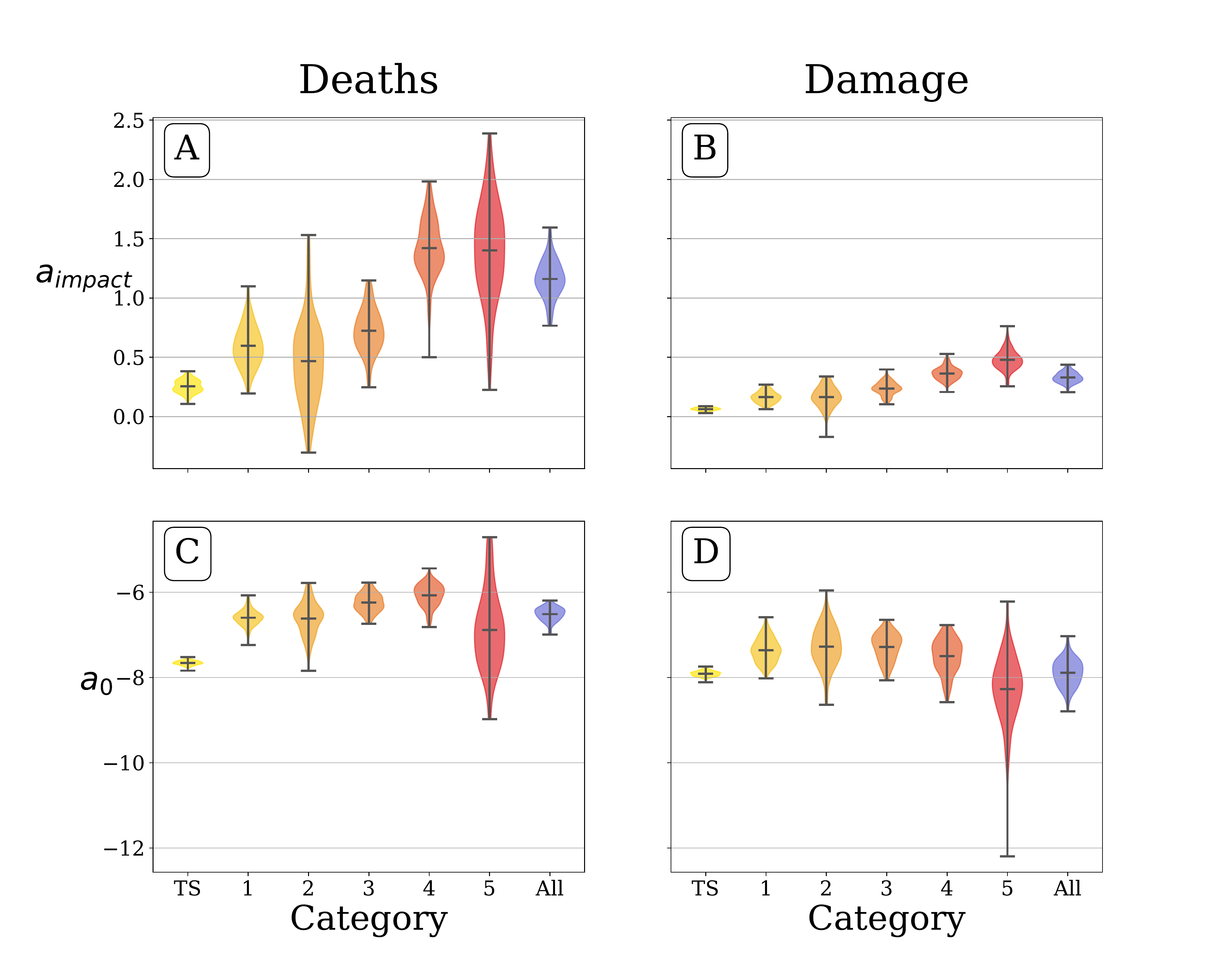}

\caption{
\textbf{Posterior distributions of regression parameters} for the model 
$\log_{10} \intUsage \sim a_0 + a_1X_{i}, $where $X_i$ is either the log number of deaths (A and C) or log damages in dollars associated with the storm (B and D), and $\log_{10} \intUsage$ is the log integrated hashtag usage rate. 
The trend in regression coefficients for association between the log attention and log deaths suggests that higher category storms receive more attention per unit impact, while the trend of intercepts shows increasing baseline attention for a hypothetical minimally disruptive storm causing exactly \$1 in damages or one death. 
For regression coefficients relating log attention to log damages, Category 4 and 5 storms receive more attention per unit increase in log damages than lower category storms. 
However, the coefficients are smaller in magnitude due to damages varying across 7 orders of magnitude, as compared to deaths varying over 4 orders of magnitude. 
There is a larger uncertainty for the category 5 intercept values, as only 6 storms of this intensity formed between 2009 and 2019 in the Atlantic basin. 
At the right of each plot, we show the coefficients for the model fit for all hurricanes (blue violin), excluding tropical storms. 
Above each category, we show the value of the mean posterior distribution for each parameter. For a table of mean parameter values, see \cref{tab:cat_parameters_mean}.
}

\label{fig:model_coefficients}
\end{figure*}

 In \cref{fig:model_coefficients} we provide posterior distributions for model parameters, which show that, as expected, more intense storms receive more attention per unit of $\log$ impact than weaker storms. 
For category five storms, we find a mean regression co-efficient of $a_{\textnormal{deaths}}= 1.35 \pm 0.39$, using the format $\mu \pm \sigma$ where $\mu$ is the mean and $\sigma$ is the standard deviation, while for category one storms we find a mean regression co-efficient of $a_{\textnormal{deaths}}= 0.61 \pm 0.18$.

Looking at associations between log damages and log attention we find  $a_{\textnormal{damage}} = 0.46 \pm 0.07$ for category 5 storms, while for category one storms we find $a_{\textnormal{damage}}= 0.17 \pm 0.05$.

 To interpret the regression coefficients, $a_{\textnormal{impact}}$, as representing proportional increases in attention per proportional increase in impact, we exponentiate the coefficient.
Thus, our model shows a 10-fold increase in deaths for a category 5 storm is associated with a 22-fold increase in attention, 
while for a category 1 storm the same 10-fold increase in deaths is associated with a 4-fold increase in attention.

The intercepts, $a_0$, for higher category storms tend to be larger, meaning that for a theoretical minimally disruptive storm causing exactly \$1 of damages or one death, a powerful storm would be talked about more, as shown in \cref{fig:model_coefficients}. 
We believe this trend could continue for category 5 storms, but we have observed only $n=6$ such storms for the duration of our attention dataset.
We interpret the intercepts as indications of how much attention low-impact storms receive on average.

In \cref{fig:model_coefficients}, we fit another regression model on all hurricanes examining log deaths and log attention. 
We find a 10-fold increase in deaths is associated with a 14-fold increase in attention, since the mean value of $\bar{a}_{\textnormal{deaths}} = 1.16 \pm 0.15$
For damages, coefficients tend to be lower than those for deaths: $\bar{a}_{\textnormal{damage}}= 0.31 \pm 0.05$. 
We intepret this coefficient as a 10-fold increase in damage being associated with no more than a 2-fold increase in attention.

\subsection{Regression Models for Impacts, Impact Interactions and Hurricane Category}
\label{subsec:combined_models}
In order to better understand the scaling of attention with hurricane impacts, we fit a number of models on the log transformed data.
We applied the same offsets as in the previous section to avoid non-finite log transformed data.
We exclude tropical storms, since their attention is not captured in same way as our string matching for hurricanes.

\subsubsection{Regression 1}\label{ss:reg1}
We fit the regression model, 
\begin{equation}
 \log_{10} \intUsage = a_0 + a_{\mathrm{death}} X_{\mathrm{death}} + a_{\mathrm{damage}} X_{\mathrm{damage}}
 + \eps,
 \end{equation}
where both predictors $X$ are log impacts, which we be referred to as regression 1. 
The regression coefficients can be interpreted as the increase in log attention received for every unit increase in log impact.
Likewise, the intercept can be interpreted as the expected attention for a minimally damaging storm causing one death and \$1 of damage. 
This model is distinguished from the previous section by including both log impacts in a single model, 
while not including an interaction term as later models will.

We set priors for the model as shown in  \cref{tab:priors1}. 
We chose the intercept, $a_0 \sim \dist{normal}(-8,3)$, to be centered around -8, approximately the lowest usage rate captured in our data, as we guess storms causing 1 death and \$1 worth of damage are talked about relatively little, but wish to allow a wide range of uncertainty spanning a few orders of magnitude.
We chose the priors for the regression coefficients $a_{\mathrm{death}}\sim a_{\mathrm{damage}} \sim \dist{normal}(0,1),$ to be weakly informative and centered around zero, as to not bias towards any association.
%\todo{Good wording here, though maybe you should set the precision of the intercept's normal distribution to be as discussed above.}
%\todo{I think having a variable hyper-prior for these these models reduces our ability to compare between iterations from regression 1 to 3. The intercept prior is reasonable }
We sampled the coefficients' posterior distributions using NUTS, using 8 chains with 2000 draws each, after 500 steps of burn-in \cite{hoffman2014no}. We found the model converged, with the maximum value of $\hat{R} = 1.000$.

We show the posterior distributions of model parameters for regression one in Panel A of \cref{fig:regressions}, which have a positive scaling between both deaths and damages, and the amount of attention commanded by the storm, as measured by the log hashtag usage rate. 
We intepret the mean value of $a_0 = -7.57 \pm 0.5$ for the regression constant as the expected log hashtag usage rate for a minimally destructive storm, i.e., that in English tweets, the hashtag usage rate would integrate to $10^{-7.57}$ over the season. We provide summary statistics in  \cref{tab:regression1}.

At first glance, this level of attention seems remarkably low: if occurring all in a single day, this is little more than 1 usage for every 100 million $1$-grams. 
The most devastating storms can have integrated usage rates of $\intUsage = 2.3\times10^{-3}$, five orders of magnitude more attention than our regression constant.
However, the least impactful storms affect relatively few people, while the most destructive storms significantly disrupt the lives of tens of millions, so the differences in the scale of total hashtag usage rate are not unreasonable. See \cref{tab:hurricane_compare} for measured values corresponding to each storm.

We find $a_{\mathrm{death}}\simeq 0.49$ and   $a_{\mathrm{damage}}\simeq 0.24.$
Because $10^{0.24}\simeq 1.7$, considering the results in linear space, a 10-fold increase in damages is associated with a 1.7-fold increase in hashtag usage rates, while a 10-fold increase in deaths is associated with a 3-fold increase.

\subsubsection{Regression 2}\label{ss:reg2}
For the second regression, 
an interaction term was introduced between
the log number of deaths and the log damages,
\begin{multline} \log_{10} \intUsage  = a_0 + a_{\mathrm{death}} X_{\mathrm{death}} + a_{\mathrm{damage}} X_{\mathrm{damage}}+ \\
a_{d, D}X_{\mathrm{death}}X_{\mathrm{damage}}
+ \eps.
\end{multline}
Prior distributions for the intercept and main effect coefficients are unchanged from regression 1, and we set the prior distribution for the interaction coefficient to be $a_{d,D}\sim \dist{normal}(0,1)$, a standard weakly informative prior for regression coefficients.
We used identical fitting procedures as above, and found the models converged with a maximum value of $\hat{R}=1.0001.$

Here, the intercept is largely the same as the simplest regression model.
Interpreting $a_{\mathrm{death}}$ as the conditional relationship
between log usage rate and log deaths when total damage is \$1, 
the $a_{\mathrm{death}}=0.05$ implies that for a 10-fold increase in deaths is associated with a 1.12-fold increase in hashtag usage rate, though the standard error includes zero. Similarly, $a_{\mathrm{damage}}=0.22$ implies a 10-fold increase in damage is associated with a 1.6-fold increase in hashtag usage rate.
Finally, the interaction coefficient $a_{\mathrm{d,D}}$ is small, but positive: a 10-fold increase in $X_{\mathrm{death}}X_{\mathrm{damage}}$ is associated with a 1.14-fold increase in hashtag usage rate.
Notably, the inclusion of the interaction term significantly reduces the regression coefficient associated with deaths, while the coefficient associated with damage is largely unchanged. 
This provides evidence that storms that cause a large number of deaths and damages are associated with higher volumes of attention, 
while a storm causing a large number of deaths but relatively less damage will attract much less attention for Twitter users. This leads us to believe that damages could act as a priming factor for human attention, in part explaining why deadly disasters in capital-poor countries often receive less attention than when similarly deadly storms occur in wealthy areas.
%\todo{discuss below}
%\todo{DRD: Also, this is very important and so you should talk about it: namely, this provides evidence that a damaging storm that also kills a lot of people attracts a lot of attention / mental space for Twitter users, while if many people are killed by a storm that does not cause a lot of damage (i.e., a storm that hits a highly-populated but capital-poor region), it does not attract a lot of attention. Is there some part of this result that points to capital destruction as the priming factor for human attention in natural disasters? Does this provide evidence to suggest why natural disasters in poor countries in Asia and Africa also do not get as much attention? You can get a lot of mileage out of this result.}. 
%\todo{MVA: Should we worry about some kind of false equivalency here? Just because the $a_{\textnormal{deaths}}$ coefficient is smaller than $a_{\textnormal{damage}}$ how can we compare them, since lives and dollars are not the same unit. Or is it really just the relative change of $a_{\textnormal{deaths}}$ when adding an interaction term, that leads us to make the priming claim.}
%%%%%%%%%%%%%%%%%%%%%%%%%%%%%%%%%%%%%%%%%%%%%%%%%%%%%%%%%%%%%%%%%%%%%%%%%%%%%%%%%%%%%%%%%%%%%
\subsubsection{Regression 3}\label{ss:reg3}
 To better understand the effect of hurricane category on attention, we performed a regression including this categorical variable, modeled as 
\begin{multline} \log_{10} \intUsage  = a_0 + a_{\mathrm{death}} X_{\mathrm{death}} + a_{\mathrm{damage}} X_{\mathrm{damage}}+ \\
a_{d, D}X_{\mathrm{death}}X_{\mathrm{damage}} +\sum_j a_{C_j} X_{C_j} + \eps,
\end{multline}
where the index $j$ runs from 2 to 5.
We did not include a variable for category 1 hurricanes to avoid issues of multi-colinearity. Fitting procedures were identical to above, and we found the model converged with the max value of $\hat{R}=1.0003$.

We did not change priors for the model coefficients from above for existing parameters, and we set the coefficients for category indicator variables to a weakly informative prior, $a_{C_i}\sim \dist{normal}(0,1)$.
Since we have included our hurricane categories, the interpretation of the intercept $a_0$ is now the expected log integrated hashtag usage rate $\intUsage$ for a category one 
hurricane, which causes one death and \$1 of damage. 
%\todo{Love this!}
The value is similar to the other regression models. Effect sizes for $a_{\mathrm{damage}}$ and $a_{\mathrm{d,D}}$ are reduced in magnitude slightly compared to the preceding regression. 

\begin{figure*}[ht!]
\centering
    \includegraphics[width=0.99\linewidth,left]{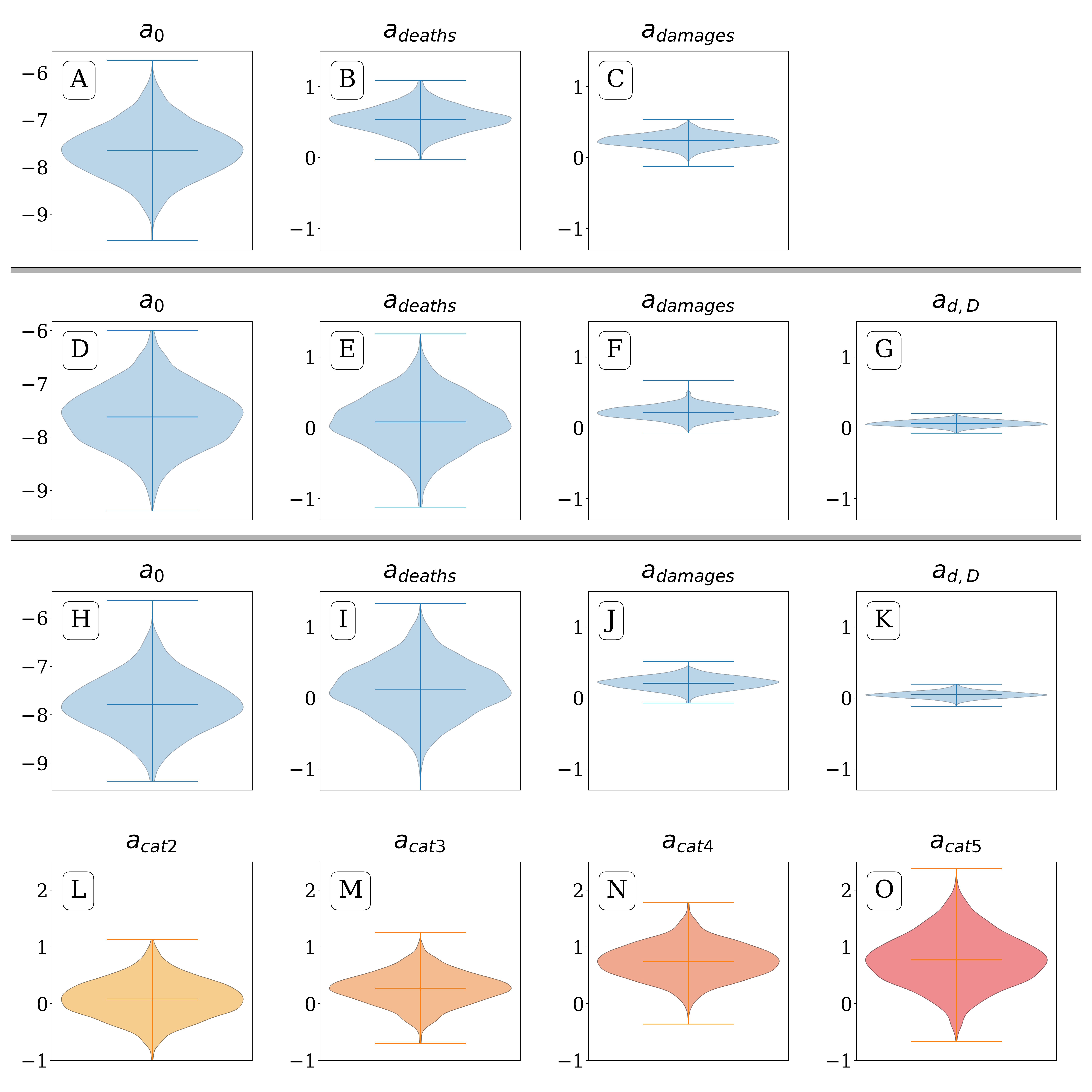}  
  \caption{
  \textbf{Parameter distributions for models 1, 2 and 3} (\cref{ss:reg1,ss:reg2,ss:reg3}). Plots A--C show posterior distributions for regression 1, $\log_{10} \intUsage \sim a_0 + a_{\textnormal{deaths}}X_{\textnormal{deaths}}+a_{\textnormal{damage}}X_{\textnormal{damage}}$, plots D--G show distributions for regression 2,  $\log_{10} \intUsage \sim a_0 + a_{\textnormal{deaths}}X_{\textnormal{deaths}}+a_{\textnormal{damage}}X_{\textnormal{damage}} + a_{d,D} X_{d,D},$ which includes the addition of an interaction term, and plots H--O showing distribution for regression 3, $\log_{10} \intUsage \sim a_0 + a_{\textnormal{deaths}}X_{\textnormal{deaths}}+a_{\textnormal{damage}}X_{\textnormal{damage}} + a_{d,D} X_{d,D} + \sum_{j=2}^5 a_{\text{cat }j}X_{\text{cat }j},$ which includes indicators variables for hurricane categories two through five. The addition of the interaction term, $a_{d,D}$ increases posterior variance for $a_{\textnormal{deaths}}$ as well as reducing its mean from $a_{\textnormal{deaths}}=0.49$ in regression 1 to $a_{\textnormal{deaths}}=0.05$ in regression 2 and $a_{\textnormal{deaths}}=0.12$ in regression 3, suggesting that while the number of deaths is associated with increased attention, attention response is primed by destruction. Additionally, the hurricane category indicator variables in regression 3 show the progressive increase in attention given to higher category storms compared to category 1 hurricanes. 
  }
  \label{fig:regressions}

\end{figure*}

 As measured by the integrated hashtag usage rate, compared to a category 1 storm causing the same deaths and damages, hurricanes in: \begin{itemize}
\item category 2 receive 1.14 times more attention,
\item category 3 receive 1.5 times more attention,
\item category 4 receive 5.6 times more attention,
\item and category 5 receive 4.6 times more attention. 
\end{itemize}
We show the posterior distributions for regression three in Panel C of \cref{fig:regressions}.

\section{Concluding Remarks}\label{sec:discussion}
We have explored the attention given to hurricanes as measured by the hashtag and 2-gram usage rate. We quantify the relative volume of attention time series for major storms. We find evidence that not only are more powerful---higher maximum category rating---storms talked about more than weaker storms, but they are talked about more when they inflict the same amount of damage or take the same number of lives.
Further, different attention scaling relationships exist for different category storms. For the most destructive storms, we demonstrate that a 10-fold increase in deaths is associated with a 27-fold increase in attention, while for weaker storms the same proportional increase in deaths would lead to only a 3-fold increase in attention on average.

How people outside of the government agencies and non-governmental organizations (NGOs) tasked with responding to natural disasters perceive the importance of disasters have real-world consequences~\cite{miller2007collective,burnside2007impact}.
We hypothesize that monetary donations to NGOs that assist with hurricane disaster relief efforts are strongly associated with the amount of attention attracted by the hurricane.
If this is true, it could be advantageous for NGOs to prospect for financial contributions while collective attention is focused most strongly on a storm~\cite{Halloran2018Sep}.
It is also possible that the speed and scale of governmental relief programs are influenced by popular attention paid to storms, and previous work has shown that relief has been inequitable in the past~\cite{Willison:2019gz}. Future work could compare the quantities of non-profit and governmental assistance with attention volume.

% defense of measuring hashtags and 2-grams
While the users of Twitter are certainly not representative of the world, or even English speakers, measuring the text they generate approaches measurement of the population at large, at least more-so than published books or edited newspaper columns~\cite{Mislove:vj, Java:2007kf, Housley:2014kh, Sloan:2015ba, Wojcik2019Apr}.
%\todo{Reviewers will get mad about this, guaranteed, so please cite like 15 things that show this}
The digital signatures left behind by our collective online presence offers rich data for observational studies of everyday language with unprecedented time resolution. Of course, many tweets referencing hurricanes are authored by journalists or news organizations and future efforts could attempt to disentangle the various motivations contributing to the overall usage rate of hashtags and other $n$-grams.

 Another limitation of our work, particularly relevant to any geospatial findings, is that we only consider tweets classified as English. While the density of English speakers closely mirrors the population density for much of the United States, we observe much lower usage rates for the English language hashtags and $2$-grams over predominately Spanish speaking areas. While different populations may use different $n$-grams to reference the same storm, for the purposes of our study we have focused only on the English-speaking population of Twitter.

%Future work
Future work could consider how to better quantify the total fraction of conversation of Twitter focused on a storm or event of interest. Our current method only includes counts for individual $n$-grams, which we believe acts as a proxy of total attention, but almost certainly underestimates the total fraction of text devoted to discussing a topic. 
Hashtag co-occurrence network-based methods could help to identify the most prominent hashtags associated with a given storm, or any event of interest, and to classify tweets as relevant. 
Examining properties of this network changing in time, such at the integrated usage rate of all significant hashtags within one degree could give a more unbiased view of the total attention surrounding the hurricane than our current method.
Other dynamics of hurricanes could be explored in this way, perhaps by encoding Jenson-Shannon Divergence shifts between hashtags as a node attribute~\cite{Dodds2020}, or more simply how the most frequently used hashtags in this ego network change in rank over time, as different phases of the storm occur.
Authors of previous works studying the effectiveness of NGO hashtag usage following natural disasters could exploit these network based methods~\cite{Wukich:2013bea}.

\acknowledgments 
The authors are grateful for the computing resources provided by the Vermont Advanced Computing Core and financial support from the Massachusetts Mutual Life Insurance Company and Google. 

\bibliography{references}
%%%%%%%%%%%%%%%%%%%%%%%%%%%%%%%%%%%%%%%%%%%%%%%%%%%%%%%%%%%

\clearpage

%% supplementary

%% following records the starting page number of the supplementary
%% section in a file called startsupp.txt
%% enables script-based breaking of manuscript and supplementary
\newwrite\tempfile
\immediate\openout\tempfile=startsupp.txt
\immediate\write\tempfile{\thepage}
\immediate\closeout\tempfile

\setcounter{page}{1}
\renewcommand{\thepage}{S\arabic{page}}
\renewcommand{\thefigure}{S\arabic{figure}}
\renewcommand{\thetable}{S\arabic{table}}
\renewcommand{\thesection}{S\arabic{section}}
\renewcommand{\thesubsection}{S\arabic{subsection}}
\setcounter{figure}{0}
\setcounter{table}{0}
\setcounter{section}{0} `

\section*{Supplementary Material}

\section{Summary Tables for Regressions}
\begin{table*}
\begin{center} 
Mean Regression Parameters -- Deaths
\end{center}
\begin{tabular}{ c c c c c c c c }
%\caption{Mean Regression Parameters -- Deaths}
\hline
 {} & Tropical Storms \hspace{1pt} & Cat 1 \hspace{1pt} & Cat 2 \hspace{1pt} & Cat 3 \hspace{1pt} & Cat 4 \hspace{1pt} & Cat 5 \hspace{1pt} & All Hurricanes \\
  \hline
  $a_{\textnormal{deaths}}$ & 0.25 & 0.61 & 0.31 & 0.72 & 1.39 & 1.35 & 1.16 \\ 
  %$10^{a_{\textnormal{deaths}}}$ & 2.69 & 2.09 & 2.5 & 3.55 & 5.62 & 26.9 & 10.47 \\
  $a_0$ & -7.65 & -6.63 & -6.58 & -6.25 & -6.01 & -6.91 & -6.56  \\ 
  \hline
\end{tabular}
    \begin{center} 
    Mean Regression Parameters -- Damages
    \end{center}
\begin{tabular}{ c c c c c c c c }
%caption{Mean Regression Parameters -- Damages}
\hline
  {} & Tropical Storms \hspace{1pt} & Cat 1 \hspace{1pt} & Cat 2 \hspace{1pt} & Cat 3 \hspace{1pt} & Cat 4 \hspace{1pt} & Cat 5 \hspace{1pt} & All Hurricanes \\
  \hline
  $a_{\textnormal{damage}}$ & 0.06 & 0.17 & 0.17 & 0.24 & 0.37 & 0.46 & 0.31 \\ 
 % $10^{a_{\textnormal{damage}}}$ & 1.20 & 1.23 & 1.55 & 1.51 & 1.51 & 3.01 & 1.91\\
  $a_0$ & -7.91 & -7.41 & -7.27 & -7.21 & -7.60 & -8.22 & -7.92  \\
  \hline
\end{tabular}
\caption{Mean Regression Parameters fit for storms of each category. See \cref{fig:model_coefficients} for full parameter distributions.} \label{tab:cat_parameters_mean}
\end{table*}

Provided for the reader here are tables of summary statistics of the estimated parameters in the regression models in \cref{subsec:category_models} and \cref{subsec:combined_models}.

%%%% 
\begin{table*}[h!]
\begin{center}
\begin{tabular}{ c c c }
\hline
 \rowcolor{gray!50}

  $a_0 $&$a_{\mathrm{death}}$&$a_{\mathrm{damage}}$\\
  \hline
  $\dist{normal}(-8,3)$ &   $\dist{normal}(0,1)$  &  $\dist{normal}(0,1)$  \\ 
  \hline
\end{tabular}
\caption{Priors for Regression 1}
\end{center}
\label{tab:priors1}

\begin{tabular}{lccccccc}
\toprule
 \rowcolor{gray!50}
 \hline
{} &  mean \hspace{1pt} &    sd\hspace{1pt} &  mc\_error \hspace{1pt}&  hpd\_2.5\hspace{1pt} &  hpd\_97.5\hspace{1pt} &    n\_eff \hspace{1pt}&  Rhat \\
\midrule
$a_0$     & -7.57 &  0.52 &      0.01 &    -8.60 &     -6.56 &  4182 &   1.0 \\
Deaths &  0.49 &  0.16 &      0.00 &     0.16 &      0.80 &  4660 &   1.0 \\
Damage &  0.24 &  0.08 &      0.00 &     0.08 &      0.40 &  4108 &   1.0 \\
sd     &  0.89 &  0.08 &      0.00 &     0.75 &      1.05 &  8449 &   1.0 \\
\bottomrule
\hline
\end{tabular}
\caption{Results for Regression 1 }
\label{tab:regression1}
\end{table*}
%%%%%%%%%%%%%%%%%%%%%%%%%%%%%%%%%%%%%%%%%%%%%%%%%%%%%%%%%%%%%%%%%%%%%%%%%%%%%%%%%%%%%%%%%%%%%

%%%%%%%%%%%%%%%%%%%%%%%%%%%%%%%%%%%%%%%%%%%%%%%%%%%%%%%%%%%%%%%%%%%%%%%%%%%%%%%%%%%%%%%%%%%%%
\begin{table*}[h!]
\begin{tabular}{ c c c c }
\hline
 \rowcolor{gray!50}
  $a_0$&  $a_{\mathrm{death}} $&$a_{\mathrm{damage}}$&$a_{\mathrm{d, D}}  $  \\
  \hline
   \rowcolor{gray!0}
  $ \dist{normal}(-8,3)$ &    $\dist{normal}(0,1)$  &    $\dist{normal}(0,1)$  &    $\dist{normal}(0,1)$  \\
  \hline
\end{tabular}
\caption{Priors for Regression 2}
\label{tab:priors2}

\rowcolors{2}{gray!25}{white}
\begin{tabular}{lccccccc}
\toprule
\hline
 \rowcolor{gray!50}
{} &  mean \hspace{1pt} &    sd\hspace{1pt} &  mc\_error \hspace{1pt}&  hpd\_2.5\hspace{1pt} &  hpd\_97.5\hspace{1pt} &    n\_eff \hspace{1pt}&  Rhat \\
\midrule
$a_0$          & -7.58 &  0.51 &      0.01 &    -8.58 &     -6.58 &   8085 &   1.0 \\
Deaths      &  0.05 &  0.34 &      0.00 &    -0.65 &      0.70 &   8326 &   1.0 \\
Damage      &  0.22 &  0.08 &      0.00 &     0.06 &      0.38 &   8151 &   1.0 \\
Interaction &  0.06 &  0.04 &      0.00 &    -0.02 &      0.14 &   8676 &   1.0 \\
sd          &  0.88 &  0.08 &      0.00 &     0.74 &      1.04 &  10843 &   1.0 \\
\bottomrule
\hline
\end{tabular}
\caption{Results for Regression 2}
\label{tab:regression2}
\end{table*}
%%%%%%%%%%%%%%%%%%%%%%%%%%%%%%%%%%%%%%%%%%%%%%%%%%%%%%%%%%%%%%%%%%%%%%%%%%%%%%%%%%%%%%%%%%%%%

\begin{table*}[h!]
\begin{tabular}{ c c c c c}
\hline
 \rowcolor{gray!50}
  $a_0 $    &$a_{\mathrm{death}}  $ &$a_{\mathrm{damage}}   $ &$a_{\mathrm{d\times D}}$ &$a_{C_i}  $\\
  \hline
   \rowcolor{gray!0}
  $ \dist{normal}(-8,3)$&   $ \dist{normal}(0,1)$&   $ \dist{normal}(0,1)$&   $ \dist{normal}(0,1)$&   $ \dist{normal}(0,1)$\\
  \hline
\end{tabular}
\caption{Priors for Regression 3}
\label{tab:priors3}

\begin{tabular}{l c c c c c c c}
\toprule
\hline
    \rowcolor{gray!50}
{} &  mean \hspace{1pt} &    sd\hspace{1pt} &  mc\_error \hspace{1pt}&  hpd\_2.5\hspace{1pt} &  hpd\_97.5\hspace{1pt} &    n\_eff \hspace{1pt}&  Rhat \\
\midrule
$a_0 $         & -7.64 &  0.51 &      0.01 &    -8.60 &     -6.60 &   9916 &   1.0 \\
Deaths      &  0.09 &  0.36 &      0.00 &    -0.60 &      0.81 &   9892 &   1.0 \\
Damage      &  0.20 &  0.08 &      0.00 &     0.05 &      0.35 &  10580 &   1.0 \\
Interaction &  0.05 &  0.04 &      0.00 &    -0.04 &      0.13 &  10424 &   1.0 \\
Cat2        &  0.07 &  0.31 &      0.00 &    -0.55 &      0.66 &  15415 &   1.0 \\
Cat3        &  0.21 &  0.26 &      0.00 &    -0.32 &      0.72 &  14877 &   1.0 \\
Cat4        &  0.76 &  0.28 &      0.00 &     0.20 &      1.29 &  15063 &   1.0 \\
Cat5        &  0.66 &  0.44 &      0.00 &    -0.17 &      1.57 &  13237 &   1.0 \\
sd          &  0.84 &  0.08 &      0.00 &     0.70 &      1.00 &  14240 &   1.0 \\
\bottomrule
\hline
\end{tabular}
\caption{Results for Regression 3}
\label{tab:regression3}
\end{table*}

\begin{table*}
\label{tab:hurricane_compare}
 \rowcolors{2}{gray!25}{white}
\begin{tabular}{lrrrrrr}
\toprule
    \rowcolor{gray!50}
    \hline
{} & Integrated Frequency \hspace{2pt} & Max Frequency \hspace{2pt} &  Deaths \hspace{2pt}&             Damage \hspace{2pt}&  Quantile 0.99 \hspace{2pt}&  Quantile 0.9 \hspace{2pt}\\
\midrule
2017 Harvey   &   $2.3\times10^{-3}$ & $3.5\times10^{-4}$ &     107 &  $1.2\times10^{11}$ &          126 &          14 \\
2017 Maria    &   $4.9\times10^{-4}$ & $5.0\times10^{-5}$ &    3057 &  $9.1\times10^{10}$ &          363 &         166 \\
2017 Irma     &   $1.6\times10^{-3}$ & $4.6\times10^{-4}$ &     134 &  $7.7\times10^{10}$ &           75 &          15 \\
2012 Sandy    &   $3.7\times10^{-4}$ & $1.5\times10^{-4}$ &     286 &  $6.8\times10^{10}$ &          157 &          13 \\
2018 Michael  &   $3.7\times10^{-4}$ & $1.1\times10^{-4}$ &      72 &  $2.5\times10^{10}$ &          201 &          13 \\
2018 Florence &   $9.3\times10^{-4}$ & $1.8\times10^{-4}$ &      57 &  $2.4\times10^{10}$ &           44 &          15 \\
2016 Matthew  &   $9.9\times10^{-4}$ & $2.6\times10^{-4}$ &     603 &  $1.6\times10^{10}$ &          136 &          15 \\
2011 Irene    &   $2.0\times10^{-4}$ & $8.0\times10^{-5}$ &      58 &  $1.4\times10^{10}$ &           14 &           8 \\
2019 Dorian   &   $5.7\times10^{-4}$ & $1.2\times10^{-4}$ &      70 &  $4.6\times10^{9 }$ &           36 &          12 \\
2012 Isaac    &   $2.6\times10^{-5}$ & $6.1\times10^{-6}$ &      41 &  $3.1\times10^{9 }$ &          192 &          97 \\
2010 Alex     &   $5.8\times10^{-6}$ & $2.5\times10^{-6}$ &      52 &  $1.5\times10^{9 }$ &           15 &           7 \\
2017 Nate     &   $6.3\times10^{-5}$ & $3.1\times10^{-5}$ &      48 &  $7.8\times10^{8 }$ &            8 &           5 \\
2019 Barry    &   $1.1\times10^{-5}$ & $3.8\times10^{-6}$ &       1 &  $6.0\times10^{8 }$ &            8 &           4 \\
2016 Hermine  &   $4.1\times10^{-5}$ & $1.9\times10^{-5}$ &       5 &  $5.5\times10^{8 }$ &            7 &           3 \\
2019 Lorenzo  &   $4.1\times10^{-6}$ & $1.0\times10^{-6}$ &      16 &  $3.6\times10^{8 }$ &           11 &           9 \\
2014 Gonzalo  &   $1.5\times10^{-5}$ & $6.4\times10^{-6}$ &       6 &  $3.1\times10^{8 }$ &           14 &          11 \\
2015 Joaquin  &   $3.7\times10^{-5}$ & $1.1\times10^{-5}$ &      34 &  $2.0\times10^{8 }$ &           11 &           5 \\
2017 Ophelia  &   $2.7\times10^{-5}$ & $1.2\times10^{-5}$ &       5 &  $8.7\times10^{7 }$ &           15 &           7 \\
2009 Bill     &   $1.6\times10^{-5}$ & $9.4\times10^{-6}$ &       2 &  $4.6\times10^{7 }$ &           11 &           7 \\
2010 Earl     &   $1.9\times10^{-5}$ & $4.9\times10^{-6}$ &       8 &  $4.5\times10^{7 }$ &            8 &           6 \\
2014 Arthur   &   $2.5\times10^{-5}$ & $1.3\times10^{-5}$ &       1 &  $1.6\times10^{7 }$ &            9 &           5 \\
2016 Nicole   &   $1.1\times10^{-5}$ & $5.3\times10^{-6}$ &       1 &  $1.5\times10^{7 }$ &           13 &           9 \\
2017 Katia    &   $4.0\times10^{-6}$ & $1.1\times10^{-6}$ &       3 &  $3.2\times10^{6 }$ &            7 &           4 \\
2017 Jose     &   $2.9\times10^{-5}$ & $4.7\times10^{-6}$ &       1 &  $2.8\times10^{6 }$ &           22 &          13 \\
2014 Bertha   &   $2.7\times10^{-6}$ & $1.1\times10^{-6}$ &       4 &               $0.0$ &           11 &           8 \\
2015 Danny    &   $4.0\times10^{-6}$ & $1.8\times10^{-6}$ &       0 &                NaN &            6 &           3 \\
\bottomrule
\hline
\end{tabular}
\caption{The unnormalized values associated with radar plots in \cref{sec:results}}
\end{table*}

\section{$2$-gram Attention Proportion of ``\texttt{hurricane}'' Usage Rate}
\label{sec:attention_share}

\begin{figure*}[t]
  \centering	
    \includegraphics[width=0.69\linewidth]{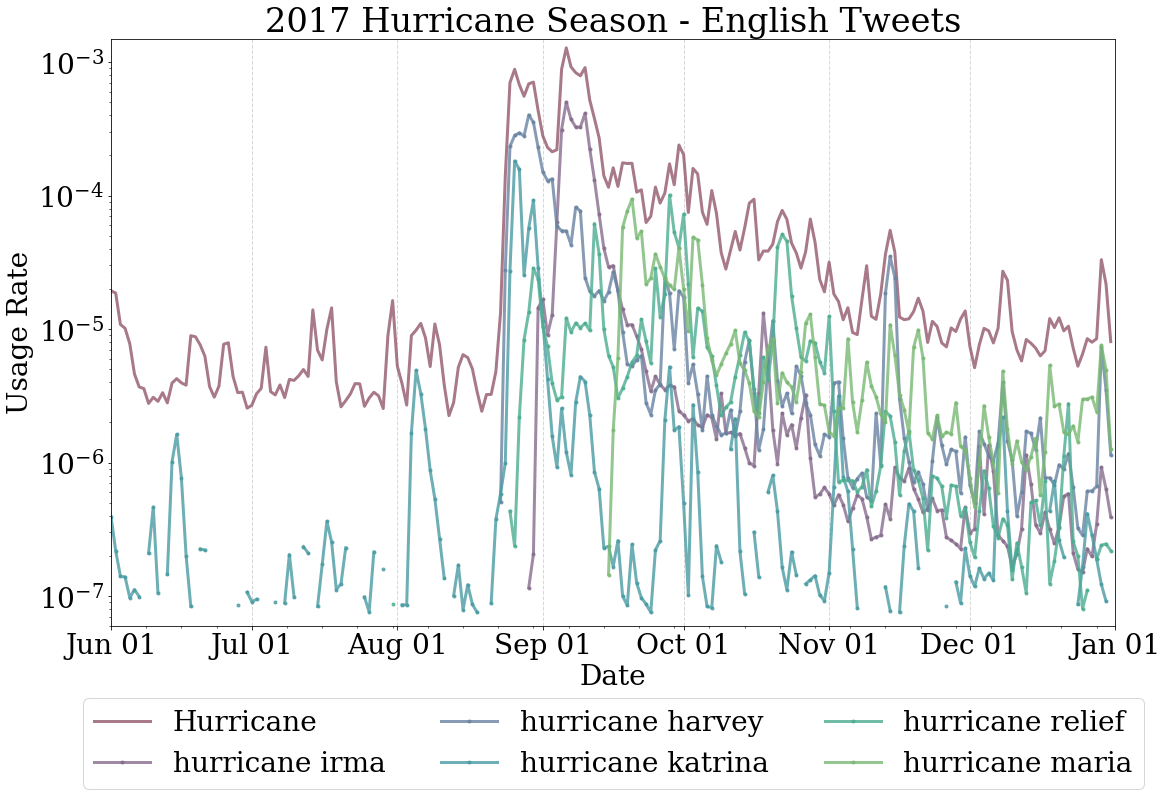} 
    \caption{Word usage rate proportions of ``\texttt{hurricane *}'' in English tweets }
\phantomsection
\label{fig:2017_attention_share_es}
    \includegraphics[width=0.69\linewidth]{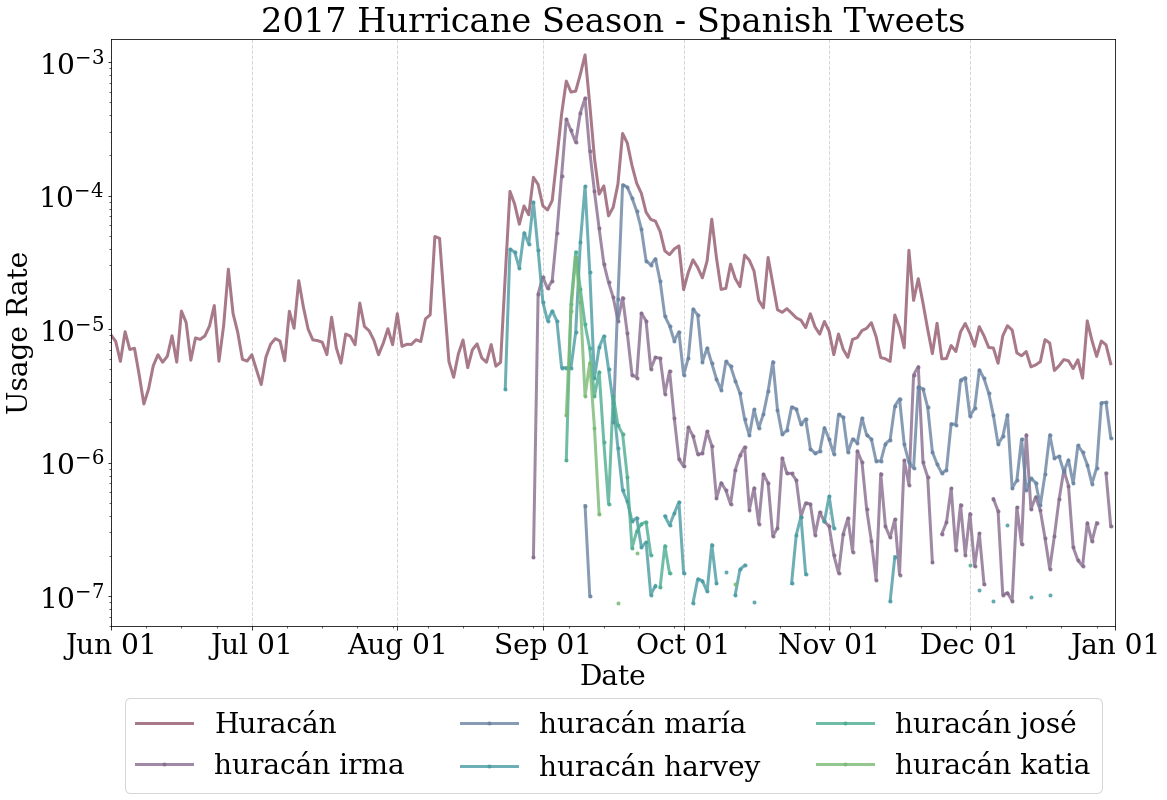}
  \caption{ Attention proportions of  ``\texttt{Huracàn *}'' in Spanish. We can see that the word usage rate surrounding ``\texttt{Hurricane Maria}'' captures a similar amount of the total attention for the $1$-gram hurricane as ``\texttt{Huracàn Marìa}'' captures. Additionally, hurricane Harvey's $2$-gram usage rate is lower in Spanish than in English, while Hurricane Katrina is talked about considerably in English but does not rise about the 50000th most used $2$-gram in Spanish. As always, usage rates are case-insensitive.}
\end{figure*}

Examining the top $2$-grams matching the pattern ``\texttt{hurricane *}'' in \cref{fig:2017_attention_share_es}, we can get a sense of what are the top storms during the season, and how much attention is allocated to each at a given time. 
For English tweets, the first major spike of the 2017 hurricane season is surrounding Hurricane Harvey, though attention also spikes for Hurricane Katrina, in reference to the 2005 storm that affected a nearby region of the gulf coast. As attention begins to decay for Hurricane Harvey, a spike in usage for the $2$-gram ``\texttt{hurricane relief}'' is observed, though it reaches only $\usageRate = 3*10^{-5}$. 
Next, attention turns to Hurricane Irma, which reaches the highest $2$-gram usage rate of any hurricane in our dataset. Finally, one week after attention for Irma begins to decay, attention spikes for Hurricane maria, though at a level noticeably lower than for Harvey or Irma.  

We notice that during storm events the $2$-gram usage rates for storms ``\texttt{hurricane *}'' is often between only half or a fifth the usage rate of the $1$-gram ``\texttt{hurricane}'',
meaning that about one in every 5 times the name of the storm follows the word hurricane in English tweets during active storms.  

In Spanish tweets the usage rates of ``\texttt{Huracàn Harvey}'' only reach a maximum of around $\usageRate\sim10^{-4}$, 
while ``\texttt{Huracàn Irma}'' receives much more relative attention.
``\texttt{Huracàn Marìa}'' receives about as much attention as Harvey, and also occupies a space similar to ``\texttt{Hurricane Maria}'' in English, around $\usageRate\sim10^{-4}$.

\section{Bi-exponential Decays}
\par
\begin{figure*}[t]
  \centering	
    \includegraphics[width=\linewidth]{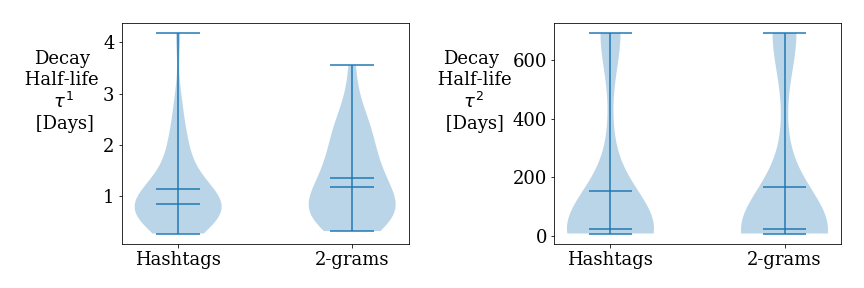}  
 \caption{\textbf{Hurricane decay half-lives:} Distributions of fitted half-lifes for the populations $u$ and $v$. The mean half-lives for $\tau_1 = 1.3$ days and $\tau_2=156$ days for hashtags and $\tau_1= 1.1$ days and $\tau_2 = 241$ days for $2$-grams. For $\tau_2$ the median half-lives are also interesting since we suspect the longest half-lives are due to poor fits. For hashtags $\tau_2 = 23$ days, and for $2$-grams $\tau_2=24$ days. }
  \label{fig:half-life_distro}
\end{figure*}

%\todo{I like this section a lot. That being said, I think that you need to include other possible models if you're going to do down the biexponential rabbit hole. The main comment that you will get from a reviewer is, well, the second part of the biexponential usually corresponds to a much longer timescale and we haven't gotten there yet. So maybe fit a power and exponential model as well and just have tables / figures comparing all of these.}
%\todo{MVA: RE: DRD Fit an exponential model, and the resulting fits are poor for storms with more than a week of data. Its clearly missing that there can be more than one decay in attention. Power laws also suck at describing the decays.}
To quantify the characteristic time scales of attention given to storms, we examined usage rates by fitting the bi-exponential model introduced by Candia \etal~\cite{Candia:2019gd}. 
Not all storms receive enough attention, but 50 of 75 in the Atlantic basin recorded at least 6 days of consecutive $2$-gram usage within the year of the hurricane, and these storms were had both their hashtag and $2$-gram usage rate fit with the bi-exponential model of Candia \etal
The model here assumes two populations, $u$ and $v$, which become interested in a given event.
Population $u$, comparable to the general population starts with a peak interest, and losses attention as $\frac{du}{dt} = -(p+r)u$. During every unit time $pu$ attention is lost from the system and $ru$ is transferred to population $v$.
The dynamics of population $v$ are as follows: $\frac{dv}{dt}= ru-qv$, so attention decays from $v$ with rate $q$, but increases proportionally to the total attention of population $u$. The final bi-exponential model is 
$$S(t) = \frac{N}{p+r-q}[(p-q)e^{-(p+r)t}+re^{-qt}],$$ and we present the half-lives associated with this model as $\tau_1 = \frac{\ln(2)}{(p+r)}$ and $\tau_2 = \frac{\ln(2)}{q}$, which are the rates of decay from the two populations $u$ and $v$.
The distributions of $\tau_1$ and $\tau_2$ for both hashtag usage rates and $2$-gram usage rates are shown in \cref{fig:half-life_distro}.
The mean half-life for population $u$, the population with faster attention decay, is $\bar{\tau_1} = 1.3$ days for hashtags, and $\bar{\tau_1} = 1.1 $ days for $2$-grams. 
The decays for population $v$ were not uni-modal, due to some storms regaining attention long after their initial impact, deviating from the model and receiveing poor fits, and resulting in very large values of $\tau_2$, but median values were approximately 24 days. 
All summary statistics are reported in \cref{tab:half-life_stats}. We speculate that for this model the population $u$ is largely people effected by the storm, while population $v$ is largely people writing about the storms or sharing information about the storm response, eg, reporters and non-profit professionals. 
Further work could look to confirm who is behind the tweets.

\begin{figure*}[t]
  \centering	
    \includegraphics[width=0.7\linewidth]{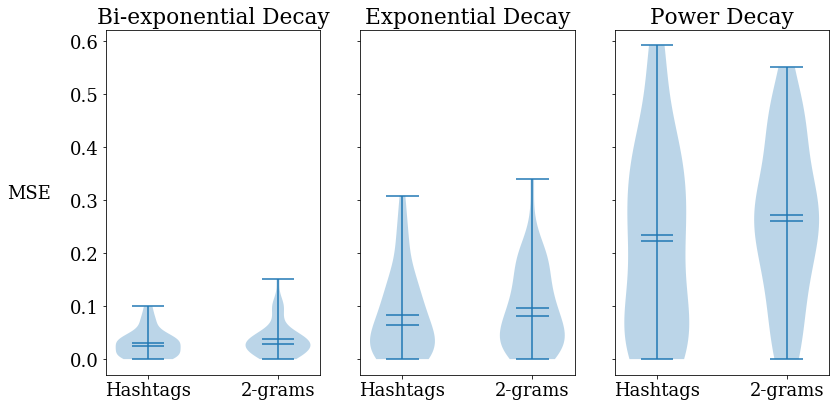}  
 \caption{\textbf{Decay Model Comparision:} Distributions of Mean Squared Error (MSE). The bi-exponential model has the lowest average MSE, followed by the simple exponential decay. The power law decay fails to capture the dynamics of attention decay, when the fit is compared to the data visually, and is reflected in the higher average MSE. }
  \label{fig:MSE}
\end{figure*}

\begin{figure*}[t]
  \centering	
    \includegraphics[width=0.31\linewidth]{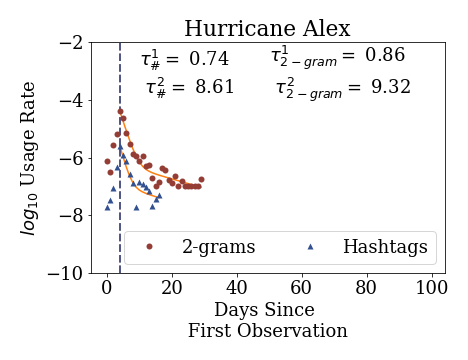}  
    \includegraphics[width=0.31\linewidth]{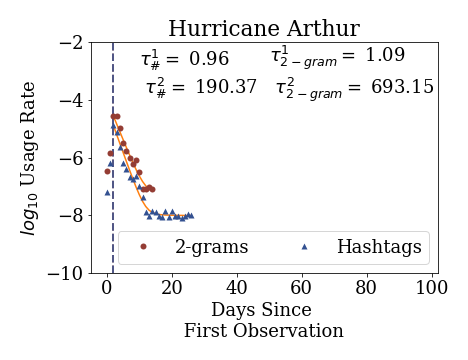} 
    \includegraphics[width=0.31\linewidth]{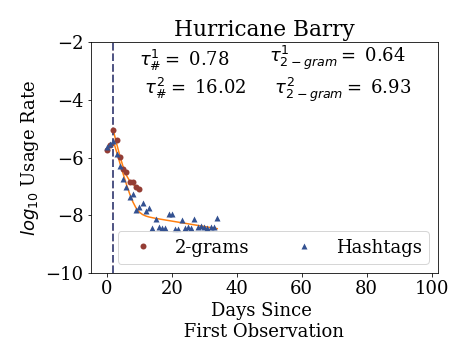}
    \includegraphics[width=0.31\linewidth]{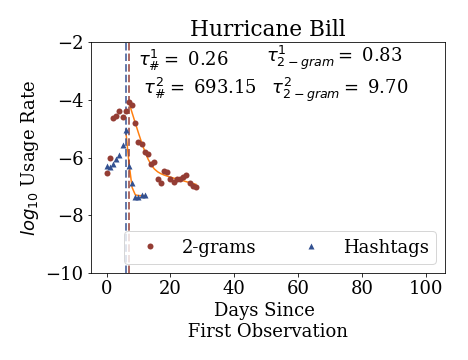}
    \includegraphics[width=0.31\linewidth]{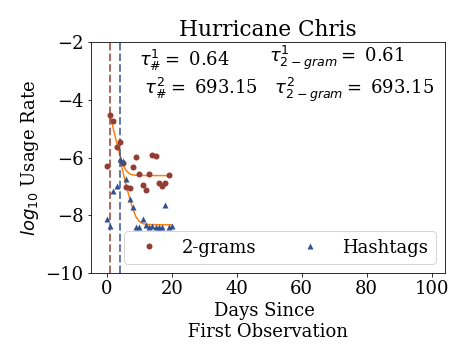}
    \includegraphics[width=0.31\linewidth]{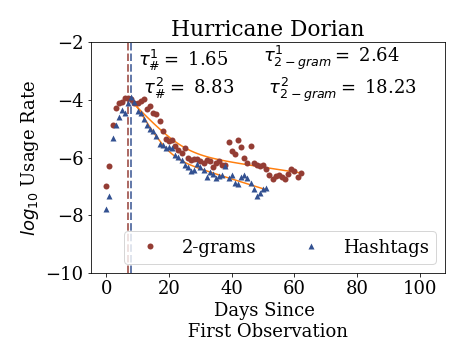}
    \includegraphics[width=0.31\linewidth]{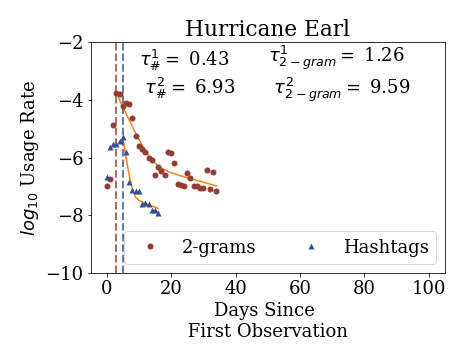}
    \includegraphics[width=0.31\linewidth]{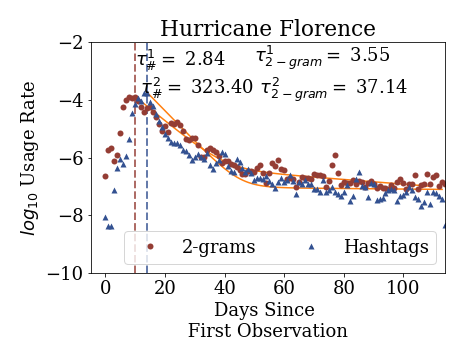}
    \includegraphics[width=0.31\linewidth]{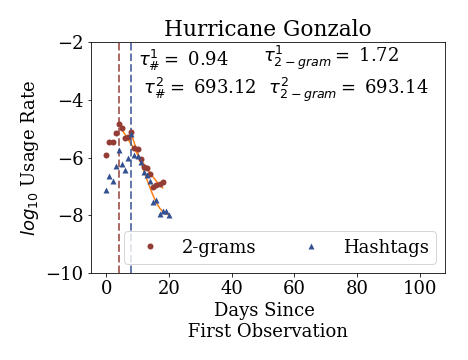}
    \includegraphics[width=0.31\linewidth]{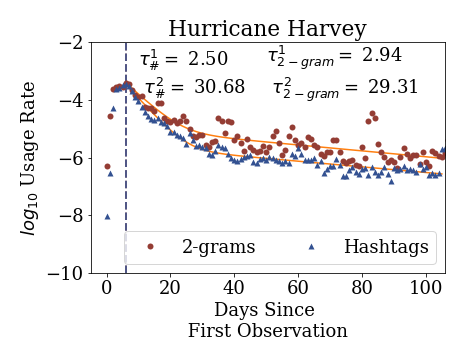}
    \includegraphics[width=0.31\linewidth]{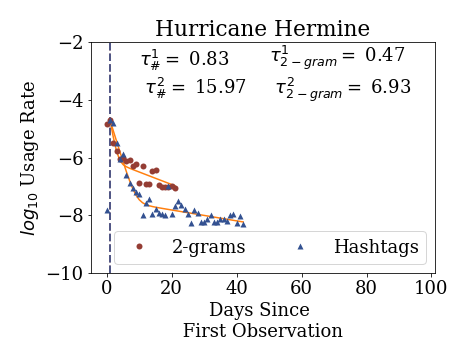}
    \includegraphics[width=0.31\linewidth]{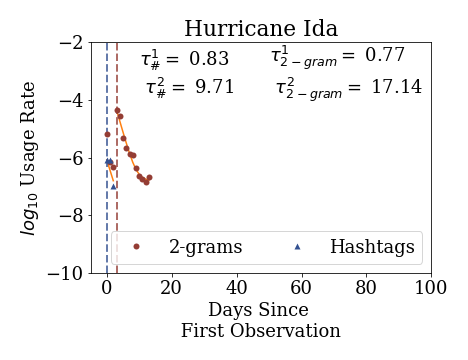}
    \includegraphics[width=0.31\linewidth]{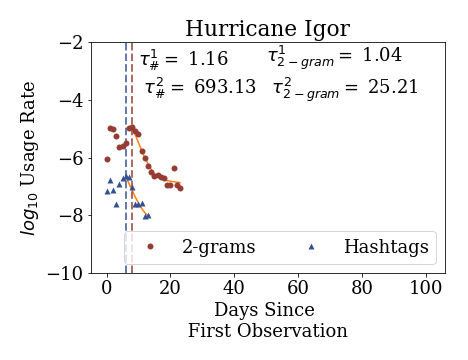}
    \includegraphics[width=0.29\linewidth]{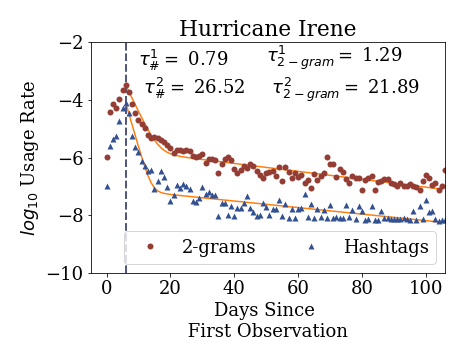}
    \includegraphics[width=0.29\linewidth]{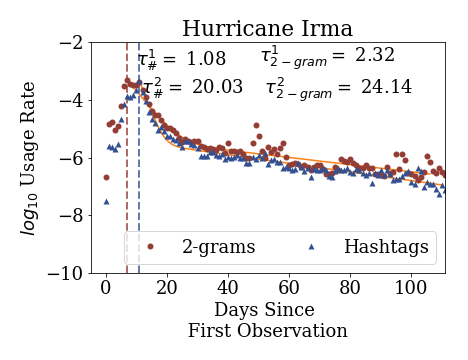}
 \caption{Hurricane bi-exponential decay fits for hashtag usage rates and $2$-gram usage rates for ``\texttt{hurricane *}''}
  \label{fig:Biexponential}
\end{figure*}

\begin{figure*}[t]
  \centering
    
    \includegraphics[width=0.29\linewidth]{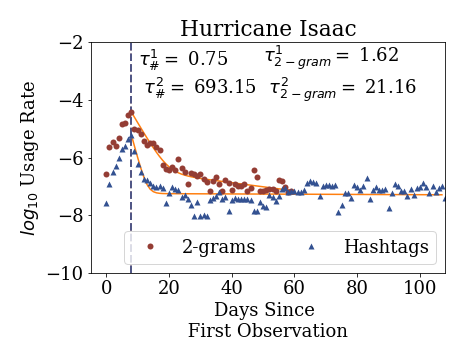}
    \includegraphics[width=0.29\linewidth]{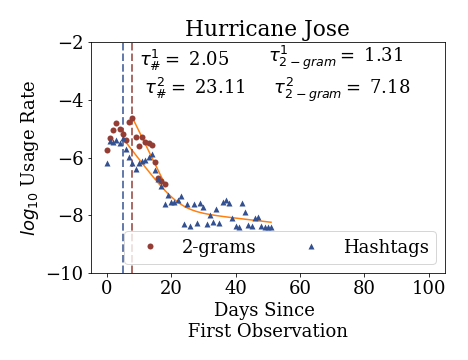}
    \includegraphics[width=0.29\linewidth]{figures/decays/decay_fit_2gramsJose.png}
    \includegraphics[width=0.29\linewidth]{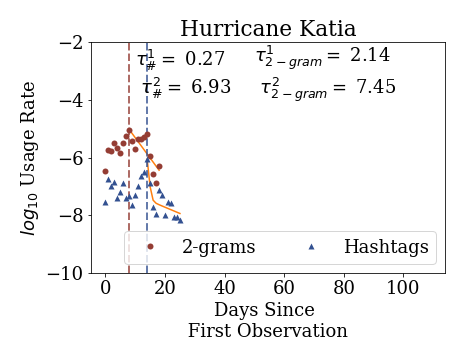}
    \includegraphics[width=0.29\linewidth]{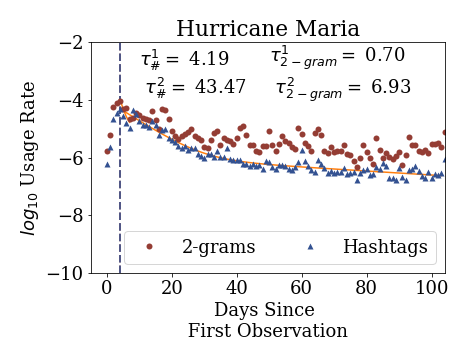}
    \includegraphics[width=0.29\linewidth]{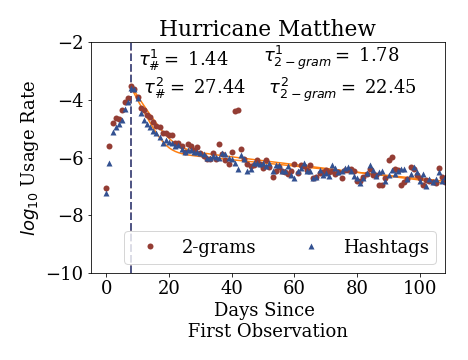}
    \includegraphics[width=0.29\linewidth]{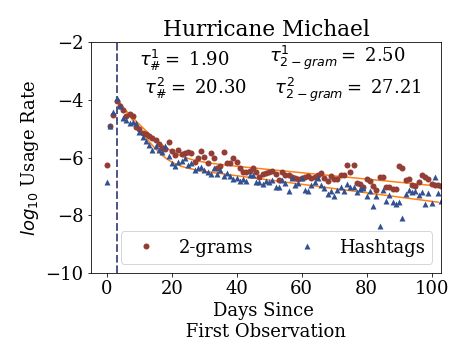}
    \includegraphics[width=0.29\linewidth]{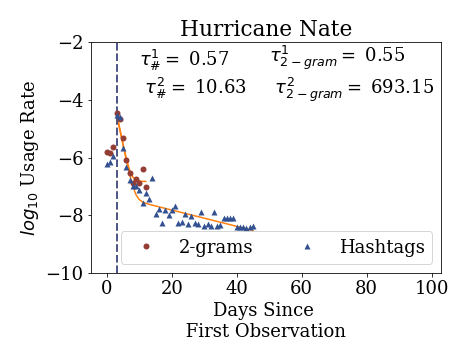}
    \includegraphics[width=0.29\linewidth]{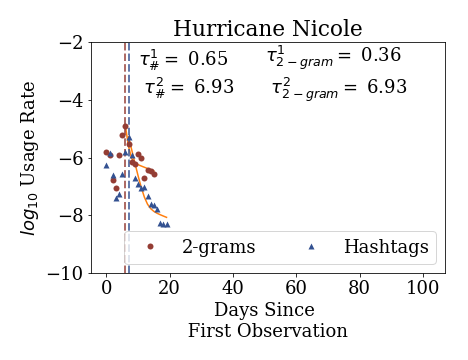}
    \includegraphics[width=0.29\linewidth]{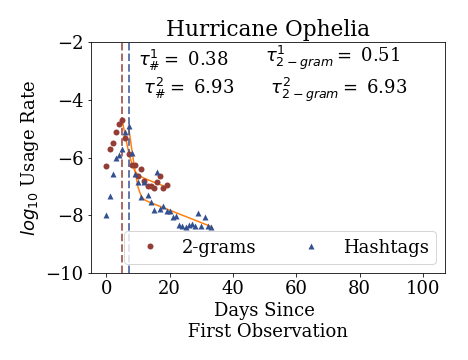}
    \includegraphics[width=0.29\linewidth]{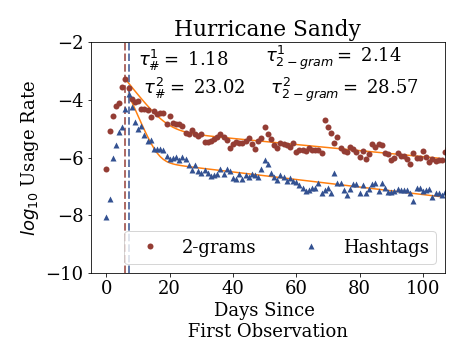}
    \includegraphics[width=0.29\linewidth]{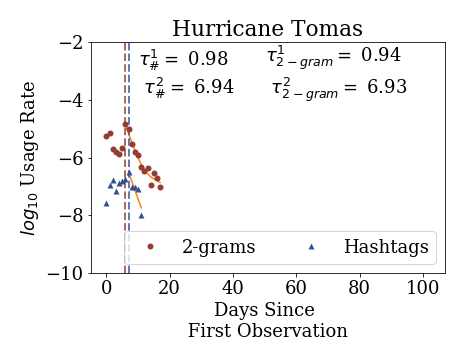}
 \caption{Hurricane decays fits for all hurricanes for which we have at least 10 days of $2$-gram usage rate data. Fits are performed for the function $y = \frac{N}{p+r-q}[(p-q)e^{-(p+r)t}+re^{-qt}]$, a simple two population decay model as proposed by Candia \etal~\cite{Candia:2019gd}. Here $p$ would be interpreted as rate of decay from population 1, $r$ would be the transfer rate from population 1 to population 2, and $r$ would be the rate of decay from population 2. Population 1 might be thought of as bystandards with a shorter attention span, while population two are those living with the ramifications, or working on the recovery who lose attention more slowly. Reported on the graph are the half lives associated with fitting this model for both the hashtag usage rate and $2$-gram usage rate, $\tau_1 = \frac{\ln2}{p+r}$ and $\tau_2 = \frac{\ln2}{q}$}
  \label{fig:biexponential2}
\end{figure*}

The fitting procedure was to first find the maximum value of the usage rate for each storm, before fitting the above model to the decay of log usage rate after this maximum. The resulting fits are shown in \cref{fig:Biexponential} and  \cref{fig:biexponential2}. The fits generally appear sensible, but there are sometimes issues for noisy time series, where the rate parameter $r$ becomes very small, corresponding to a very long half-life, and misfitting the early decay. This occurs in the time series for Hurricane Florence. The distributions of Mean Squared Error (MSE) are shown in \cref{fig:MSE}.

\begin{figure*}[t]
  \centering	
    \includegraphics[width=0.5\linewidth]{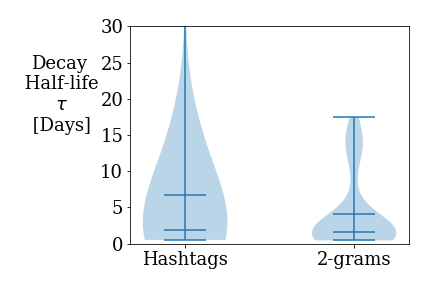}  
 \caption{\textbf{Simple Exponential Hurricane decay half-lives:} Distributions of fitted half-lives for a single population. The median half-lives for $\tau = 5.3$ days a for hashtags and $\tau= 5.2$  days for $2$-grams. The simple exponential model fails to explain the break in attention decay for larger storms, receiving more attention. The bi-modal distribution of half-lives for 2-grams suggests that there are two categories of storms, ones with larger half-lives have more data, and thus the longer decay increases the fitted half-life. Meanwhile, smaller storms receive so little attention, that we don't measure any after a week or so, leading to a much smaller half-live, which corresponds to $\tau_1$ in our bi-exponential fit. }
  \label{fig:simple_halflife}
\end{figure*}

\begin{table*}
\begin{tabular}{lrrrr}
\toprule
\hline
 \rowcolor{gray!50}
{} &           Max Usage Rate \hspace{2pt}&  $\tau_1$ [Days]\hspace{2pt} &  $\tau_2$ [Days]\hspace{2pt}&   \hspace{3pt}Season \\
\midrule
\#hurricanealex      & $2.5\times10^{-6}$ &    0.7 &    8.6 & 2010 \\
\#hurricanearthur    & $1.3\times10^{-5}$ &    0.9 &  190.3 & 2014 \\
\#hurricanebarry     & $3.8\times10^{-6}$ &    0.7 &   16.0 & 2019 \\
\#hurricanebertha    & $1.1\times10^{-6}$ &    0.6 &    6.9 & 2014 \\
\#hurricanebill      & $9.4\times10^{-6}$ &    0.2 &  693.1 & 2009 \\
\#hurricanechris     & $8.9\times10^{-7}$ &    0.6 &  693.1 & 2018 \\
\#hurricanecristobal & $2.0\times10^{-7}$ &    2.0 &    6.9 & 2014 \\
\#hurricanedanielle  & $1.9\times10^{-7}$ &    0.7 &  693.1 & 2010 \\
\#hurricanedanny     & $1.8\times10^{-6}$ &    0.7 &    6.9 & 2015 \\
\#hurricanedorian    & $1.2\times10^{-4}$ &    1.6 &    8.8 & 2019 \\
\#hurricaneearl      & $5.0\times10^{-6}$ &    0.4 &    6.9 & 2010 \\
\#hurricaneflorence  & $1.8\times10^{-4}$ &    2.8 &  323.3 & 2018 \\
\#hurricanegert      & $3.6\times10^{-7}$ &    0.4 &    6.9 & 2017 \\
\#hurricanegonzalo   & $6.4\times10^{-6}$ &    0.9 &  693.1 & 2014 \\
\#hurricaneharvey    & $3.5\times10^{-4}$ &    2.5 &   30.6 & 2017 \\
\#hurricanehermine   & $1.9\times10^{-5}$ &    0.8 &   15.9 & 2016 \\
\#hurricaneida       & $8.3\times10^{-7}$ &    0.8 &    9.7 & 2009 \\
\#hurricaneigor      & $2.2\times10^{-7}$ &    1.1 &  693.1 & 2010 \\
\#hurricaneirene     & $8.0\times10^{-5}$ &    0.7 &   26.5 & 2011 \\
\#hurricaneirma      & $4.6\times10^{-4}$ &    1.0 &   20.0 & 2017 \\
\#hurricaneisaac     & $6.1\times10^{-6}$ &    0.7 &  693.1 & 2012 \\
\#hurricanejoaquin   & $1.1\times10^{-5}$ &    1.2 &   57.7 & 2015 \\
\#hurricanejose      & $4.7\times10^{-6}$ &    2.0 &   23.1 & 2017 \\
\#hurricanekarl      & $7.4\times10^{-8}$ &    0.6 &   68.9 & 2010 \\
\#hurricanekatia     & $8.7\times10^{-7}$ &    0.2 &    6.9 & 2011 \\
\#hurricanelorenzo   & $1.0\times10^{-6}$ &    1.3 &   64.2 & 2019 \\
\#hurricanemaria     & $5.0\times10^{-5}$ &    4.1 &   43.4 & 2017 \\
\#hurricanematthew   & $2.6\times10^{-4}$ &    1.4 &   27.4 & 2016 \\
\#hurricanemichael   & $1.1\times10^{-4}$ &    1.8 &   20.2 & 2018 \\
\#hurricanenate      & $3.1\times10^{-5}$ &    0.5 &   10.6 & 2017 \\
\#hurricanenicole    & $5.3\times10^{-6}$ &    0.6 &    6.9 & 2016 \\
\#hurricaneophelia   & $1.2\times10^{-5}$ &    0.3 &    6.9 & 2017 \\
\#hurricanesandy     & $1.5\times10^{-4}$ &    1.1 &   23.0 & 2012 \\
\#hurricanetomas     & $3.0\times10^{-7}$ &    0.9 &    6.9 & 2010 \\
\bottomrule
\hline
\end{tabular}
\caption{Fitted  half-lives $\tau_1$ and $\tau_2$  for all storms with at least 10 days of hashtag usage.}
\end{table*}

\begin{table*}
\begin{tabular}{lrrrr}
\toprule
\hline
{} &           Max Usage Rate \hspace{2pt}&  $\tau_1$ [Days]\hspace{2pt} &  $\tau_2$ [Days]\hspace{2pt}&   \hspace{3pt}Season \\
\midrule
Hurricane Alex      & $4.1\times10^{-5}$ &    0.8 &    9.3 & 2010 \\
Hurricane Arthur    & $2.8\times10^{-5}$ &    1.0 &  693.1 & 2014 \\
Hurricane Barry     & $8.9\times10^{-6}$ &    0.6 &    6.9 & 2019 \\
Hurricane Bertha    & $8.2\times10^{-6}$ &    0.4 &  693.1 & 2014 \\
Hurricane Bill      & $8.2\times10^{-5}$ &    0.8 &    9.7 & 2009 \\
Hurricane Chris     & $3.0\times10^{-5}$ &    0.6 &  693.1 & 2018 \\
Hurricane Cristobal & $1.9\times10^{-6}$ &    1.5 &  693.1 & 2014 \\
Hurricane Danielle  & $1.0\times10^{-5}$ &    0.9 &    7.1 & 2010 \\
Hurricane Danny     & $7.6\times10^{-6}$ &    0.6 &  693.1 & 2015 \\
Hurricane Dorian    & $1.1\times10^{-4}$ &    2.6 &   18.2 & 2019 \\
Hurricane Earl      & $1.7\times10^{-4}$ &    1.2 &    9.5 & 2010 \\
Hurricane Florence  & $1.3\times10^{-4}$ &    3.5 &   37.1 & 2018 \\
Hurricane Gert      & $1.0\times10^{-6}$ &    2.1 &  321.9 & 2017 \\
Hurricane Gonzalo   & $1.4\times10^{-5}$ &    1.7 &  693.1 & 2014 \\
Hurricane Harvey    & $4.0\times10^{-4}$ &    2.9 &   29.3 & 2017 \\
Hurricane Hermine   & $2.0\times10^{-5}$ &    0.4 &    6.9 & 2016 \\
Hurricane Ida       & $4.5\times10^{-5}$ &    0.7 &   17.1 & 2009 \\
Hurricane Igor      & $1.1\times10^{-5}$ &    1.0 &   25.2 & 2010 \\
Hurricane Irene     & $3.3\times10^{-4}$ &    1.2 &   21.8 & 2011 \\
Hurricane Irma      & $5.0\times10^{-4}$ &    2.3 &   24.1 & 2017 \\
Hurricane Isaac     & $3.8\times10^{-5}$ &    1.6 &   21.1 & 2012 \\
Hurricane Joaquin   & $4.4\times10^{-5}$ &    1.2 &  144.5 & 2015 \\
Hurricane Jose      & $2.4\times10^{-5}$ &    1.3 &    7.1 & 2017 \\
Hurricane Karl      & $1.6\times10^{-5}$ &    0.3 &    6.9 & 2010 \\
Hurricane Katia     & $9.3\times10^{-6}$ &    2.1 &    7.4 & 2011 \\
Hurricane Lorenzo   & $2.7\times10^{-6}$ &    1.7 &    8.1 & 2019 \\
Hurricane Maria     & $1.1\times10^{-4}$ &    0.7 &    6.9 & 2017 \\
Hurricane Matthew   & $2.9\times10^{-4}$ &    1.7 &   22.4 & 2016 \\
Hurricane Michael   & $9.3\times10^{-5}$ &    2.5 &   27.2 & 2018 \\
Hurricane Nate      & $3.5\times10^{-5}$ &    0.5 &  693.1 & 2017 \\
Hurricane Nicole    & $1.2\times10^{-5}$ &    0.3 &    6.9 & 2016 \\
Hurricane Ophelia   & $1.9\times10^{-5}$ &    0.5 &    6.9 & 2017 \\
Hurricane Sandy     & $5.3\times10^{-4}$ &    2.1 &   28.5 & 2012 \\
Hurricane Tomas     & $1.4\times10^{-5}$ &    0.9 &    6.9 & 2010 \\
\bottomrule
\hline
\label{tab:half-life_stats}
\end{tabular}
\caption{Fitted half-lives $\tau_1$ and $\tau_2$ for all storms with at least 10 days of $2$-gram usage.}
\end{table*}

Looking at the decay half-lives in Table \ref{tab:half-life_stats} we notice can see that most hurricane hashtags lose half their volume on the order of 1 or 2 days.
The storms with relatively more attention on Twitter, Harvey, Irma, Matthew, and Sandy, all initially decay quickly, with a half-life on the order of a few days, but then have much longer decays associated with $\tau_2$, on the order of a few weeks. 
There are some aberrations where the bi-exponential model does a poor job of explaining the data, such as for hurricane Joaquin, where a fight between Governor Bobby Jindal and the Obama administration over the size of a recovery package spurred news stories and attention long after the initial activity associated with the storm itself. 
This leads to increases in hashtag usage rate, and thus negative half-lives. 
The longest half-life is associated with hurricane Maria, $\tau_2$ was approximately twice as long as the next largest hurricane. 
The extended crisis in Puerto Rico caused by Maria may be a reason this exceedingly long lifetime, even though the initial attention received by the hashtag was less than storms of comparable strength. 

We also fit a simple exponential model $S(t) = N e^{-pt}.$ For high attention storms for which we have more than a week of data, this model is unable to capture decays occurring on different time scales, and thus has poor fits. For smaller storms for which attention is lower than the resolution of our data set, the exponential model is perhaps more appropriate. A distribution of half-lives for hashtags and $2$-grams is shown in \cref{fig:simple_halflife}. While for larger storms, the fits did not capture the changing rates of attention decay, it was adequate for smaller storms that decay quickly below our instrument's resolution. However, for storms for which we have data for an extended decay, the bi-exponential model is more appropriate. 

\section{Hurricane Attention Maps}
The remaining Hurricane Attention Map and time series from 2009 to 2018 are presented for the reader's perusal. Only storms reaching at least Category 2 are shown, and Seasons 2013 and 2014 are omitted. Earlier storms in our dataset mostly did not make landfall, and thus appear to recieve relatively little attention. The scale of attention on the maps is held constant between years. 

\begin{figure*}[t]
  \centering	
    \includegraphics[width=\linewidth]{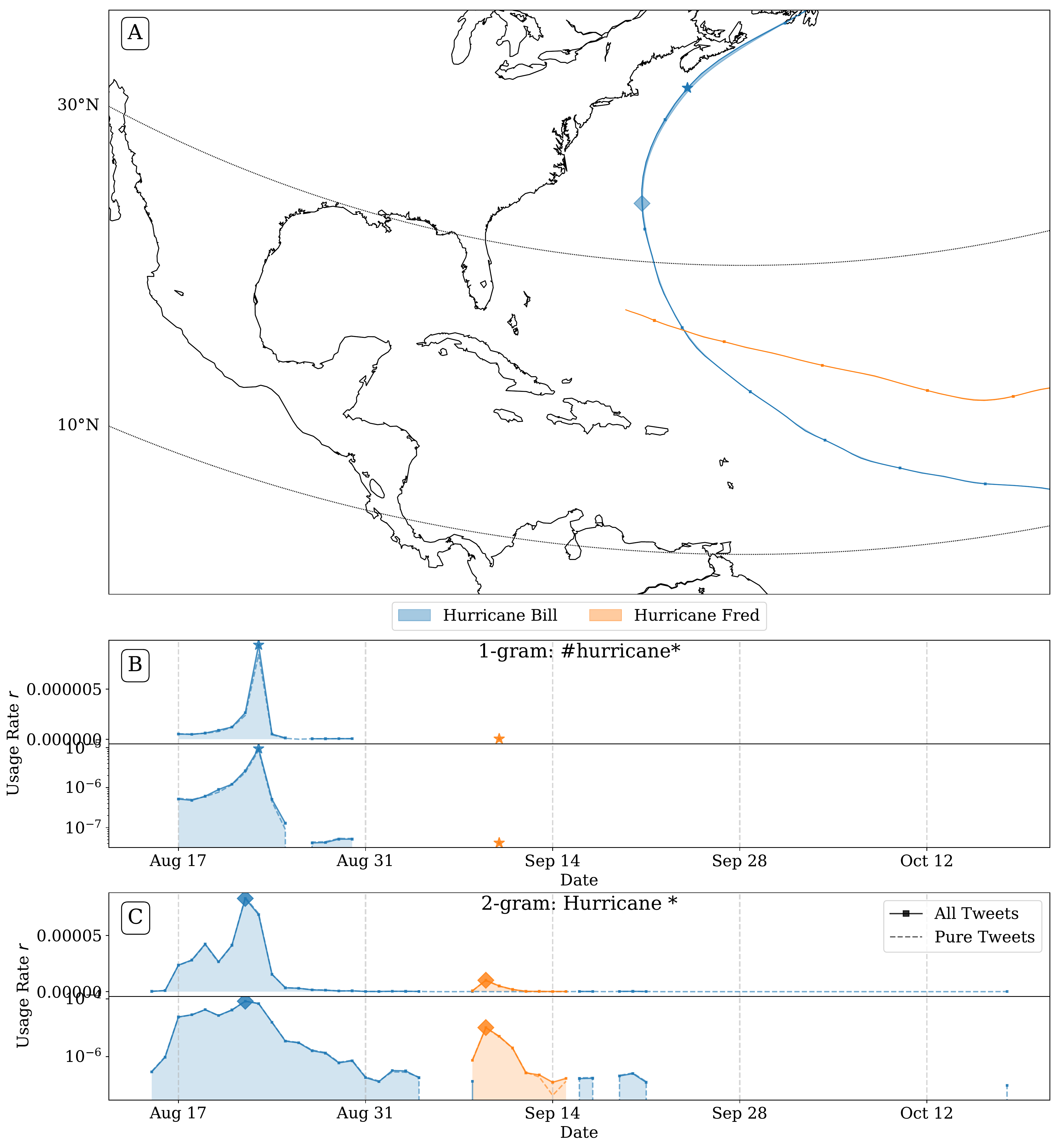}  
 \caption{Hurricane Attention Map and time series for 2009}
  \label{fig:2009}
\end{figure*}

\begin{figure*}[ht]
  \centering	
    \includegraphics[width=\linewidth]{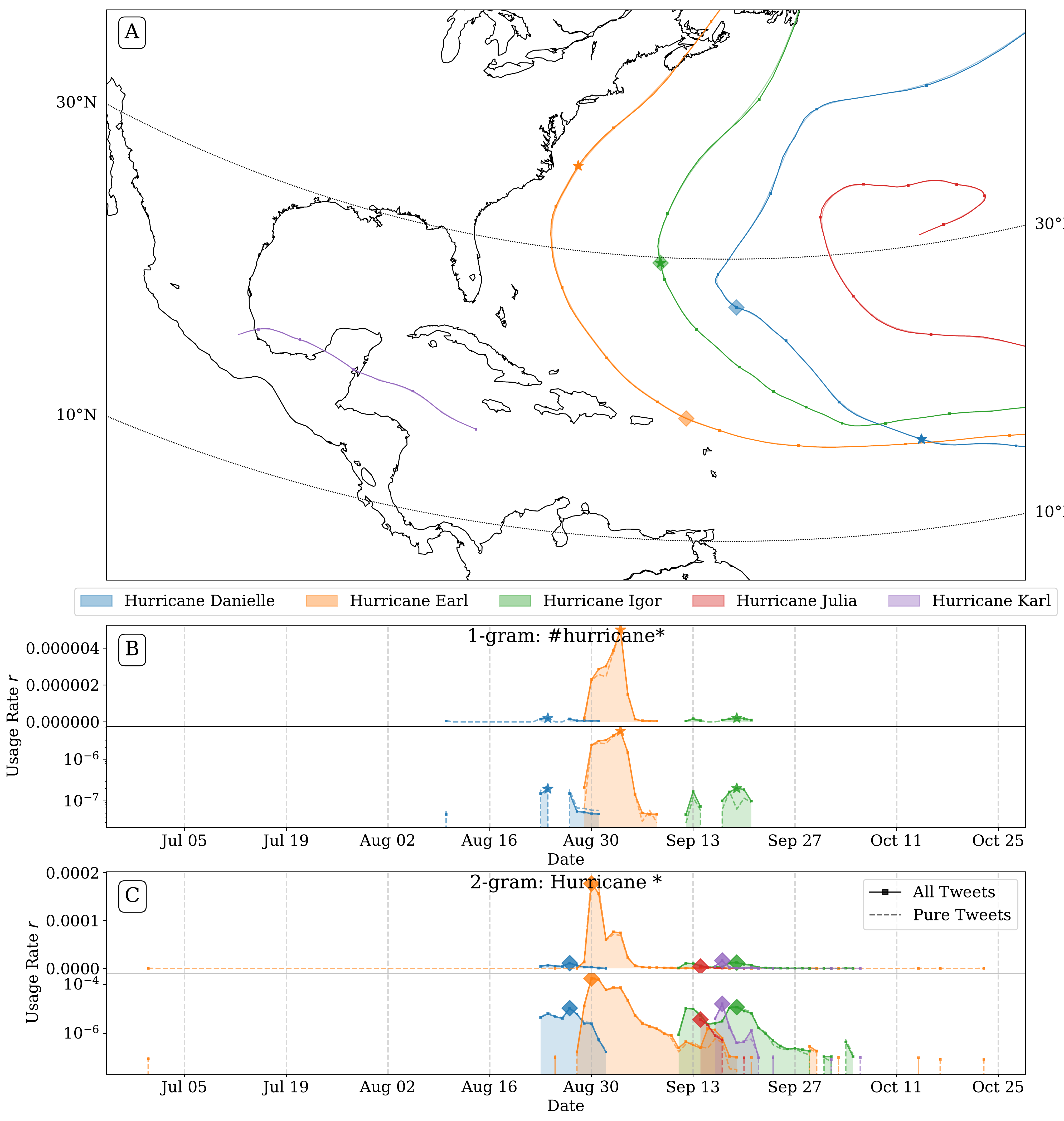}  
 \caption{Hurricane Attention Map and time series for 2010}
  \label{fig:2010}
\end{figure*}

\begin{figure*}[ht]
  \centering	
    \includegraphics[width=\linewidth]{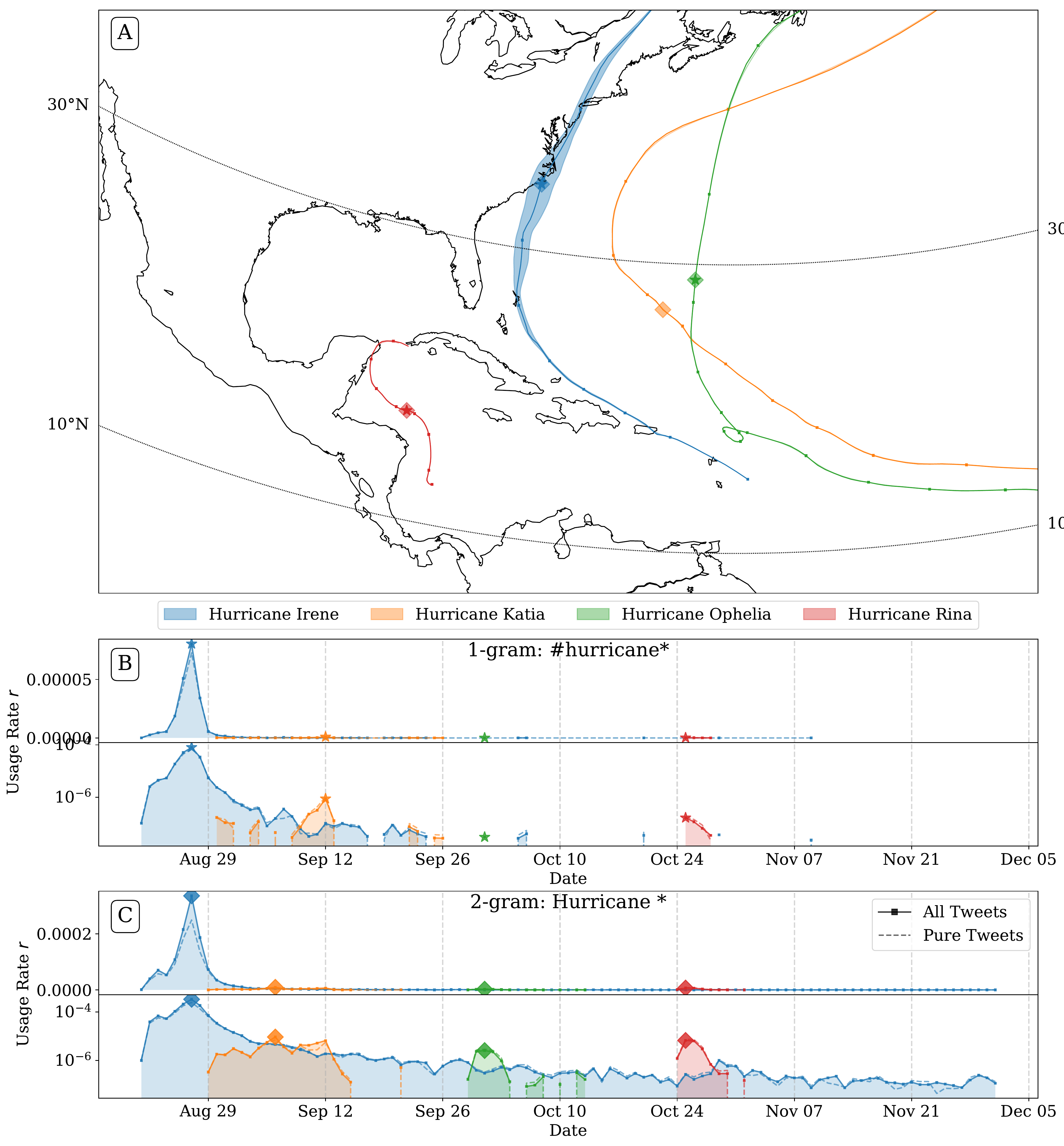}  
 \caption{Hurricane Attention Map and time series for 2011}
  \label{fig:2011}
\end{figure*}

\begin{figure*}[ht]
  \centering	
    \includegraphics[width=0.9\linewidth]{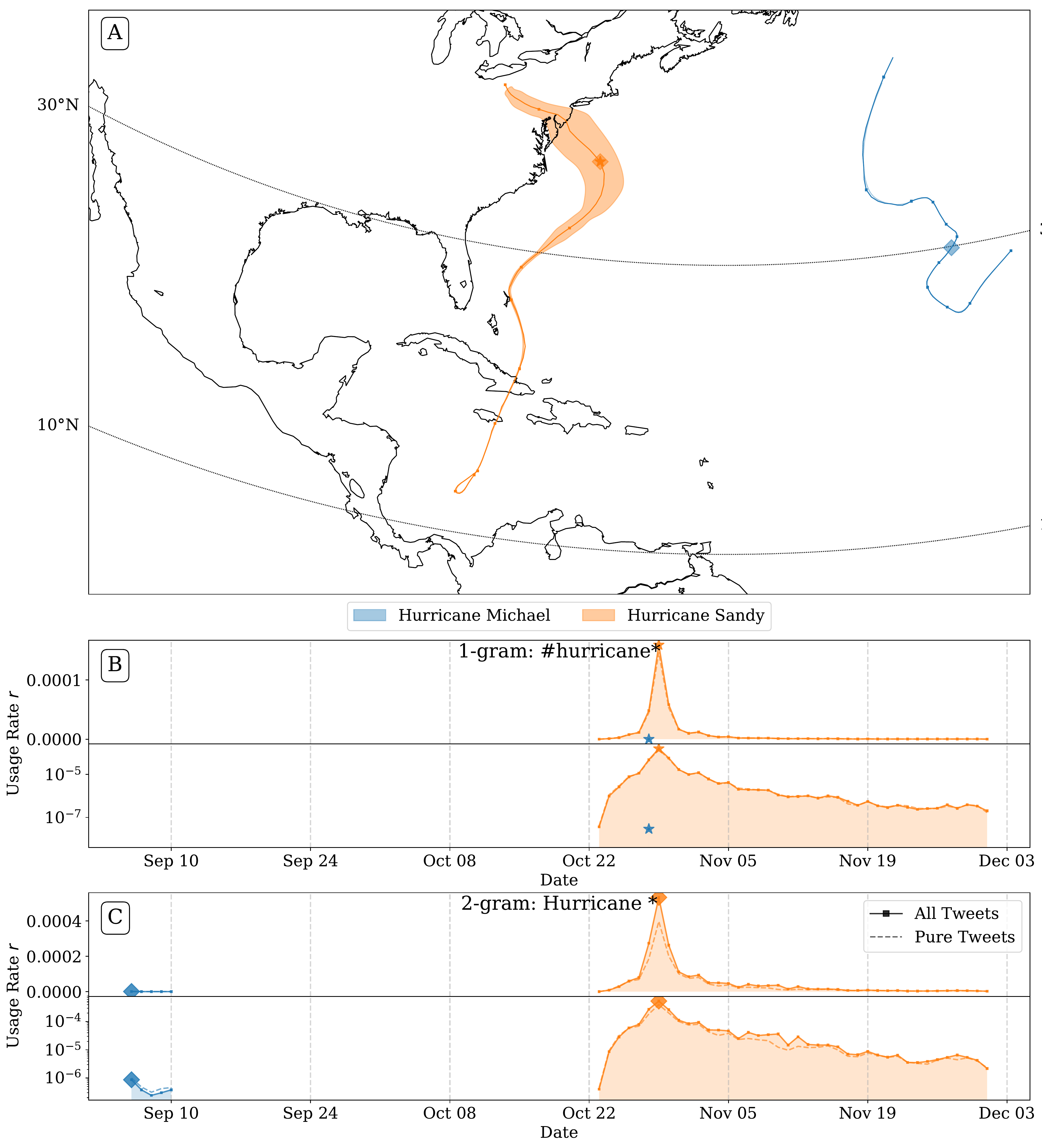}  
 \caption{Hurricane Attention Map and time series for 2012}
  \label{fig:2012}
\end{figure*}

\begin{figure*}[ht]
  \centering	
    \includegraphics[width=0.9\linewidth]{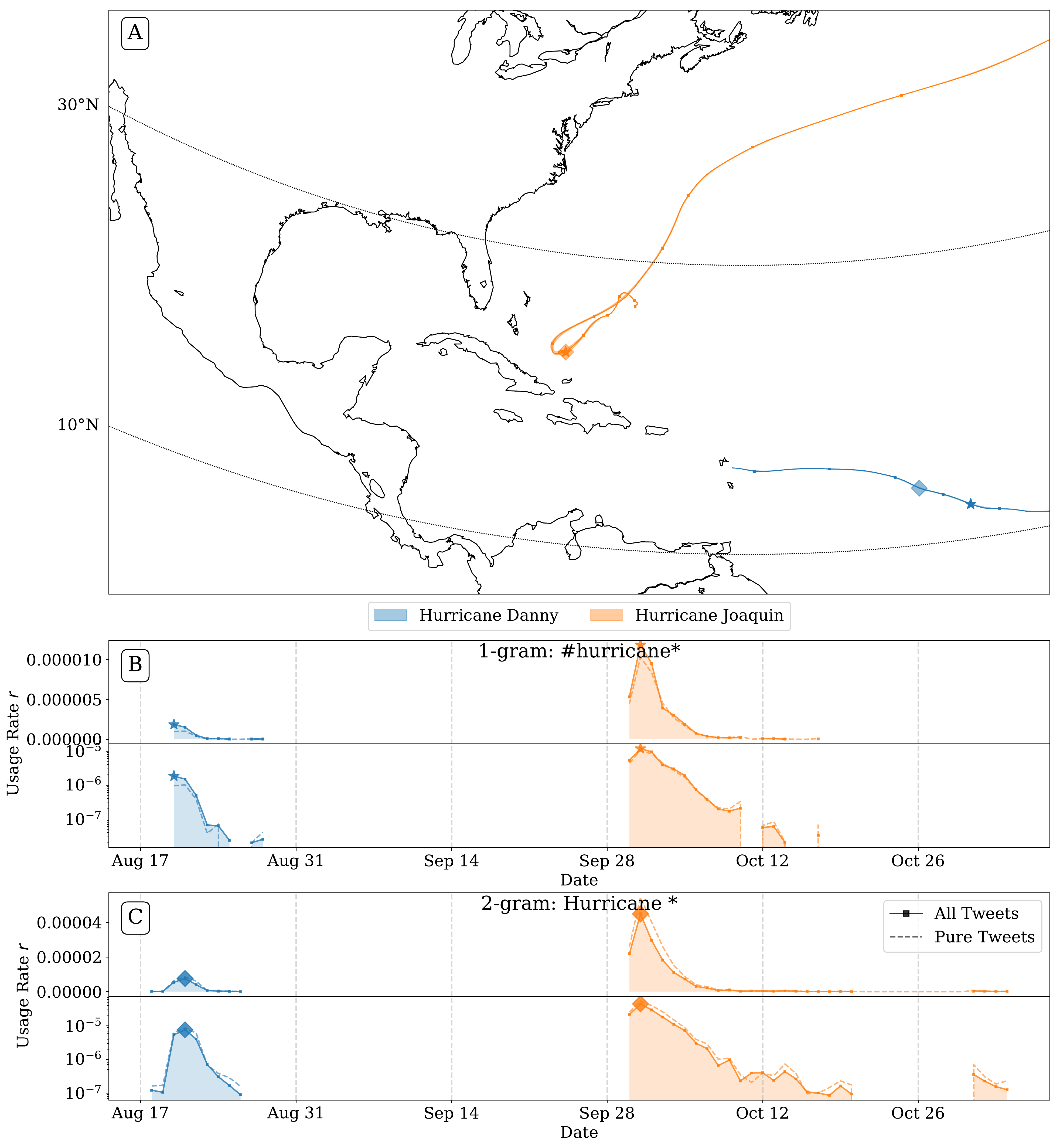}  
 \caption{Hurricane Attention Map and time series for 2015}
  \label{fig:2015}
\end{figure*}

\begin{figure*}[ht]
  \centering	
    \includegraphics[width=0.9\linewidth]{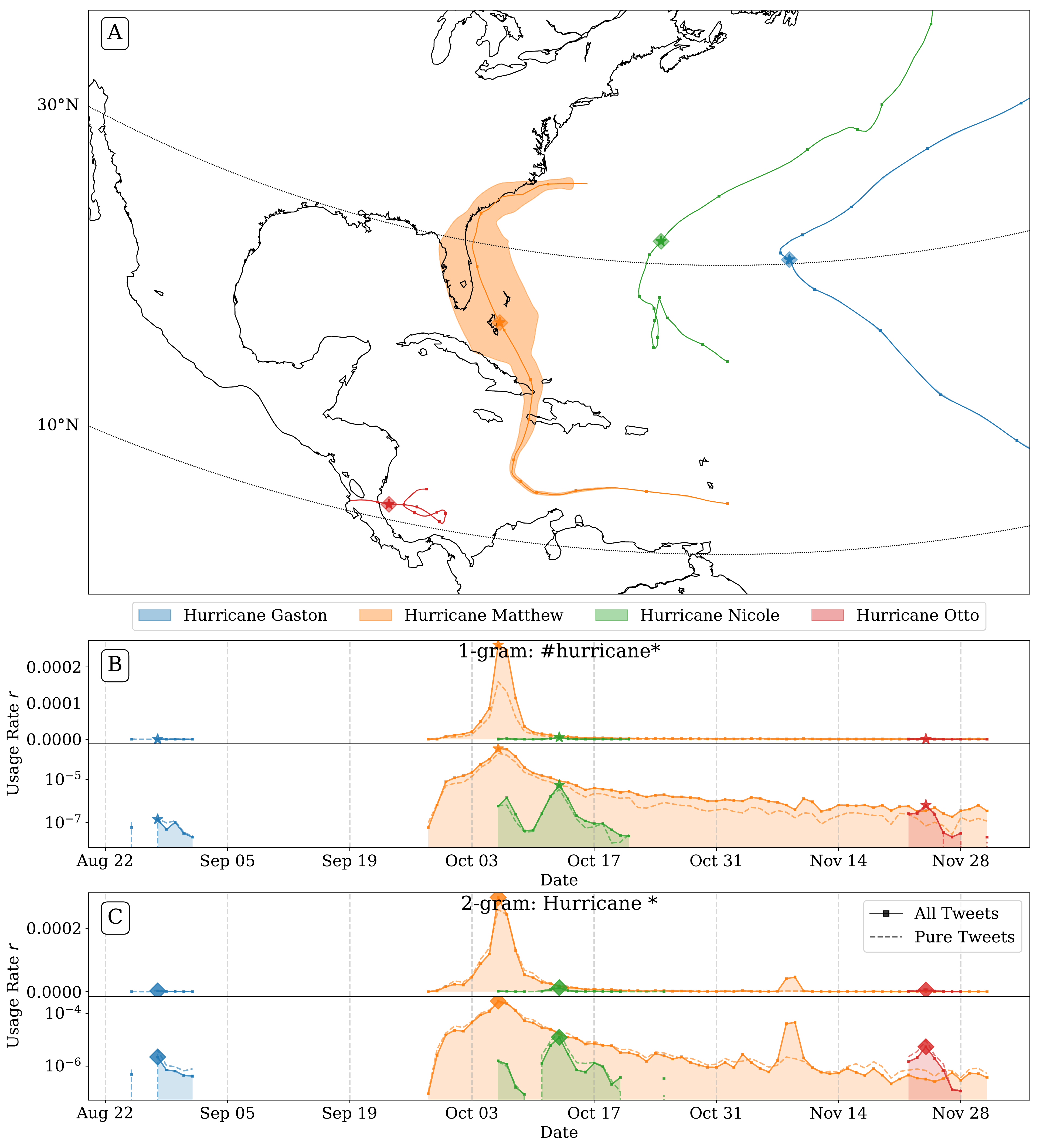}  
 \caption{Hurricane Attention Map and time series for 2016}
  \label{fig:2016}
\end{figure*}

\begin{figure*}[ht]
\centering	
\includegraphics[width=0.9\linewidth]{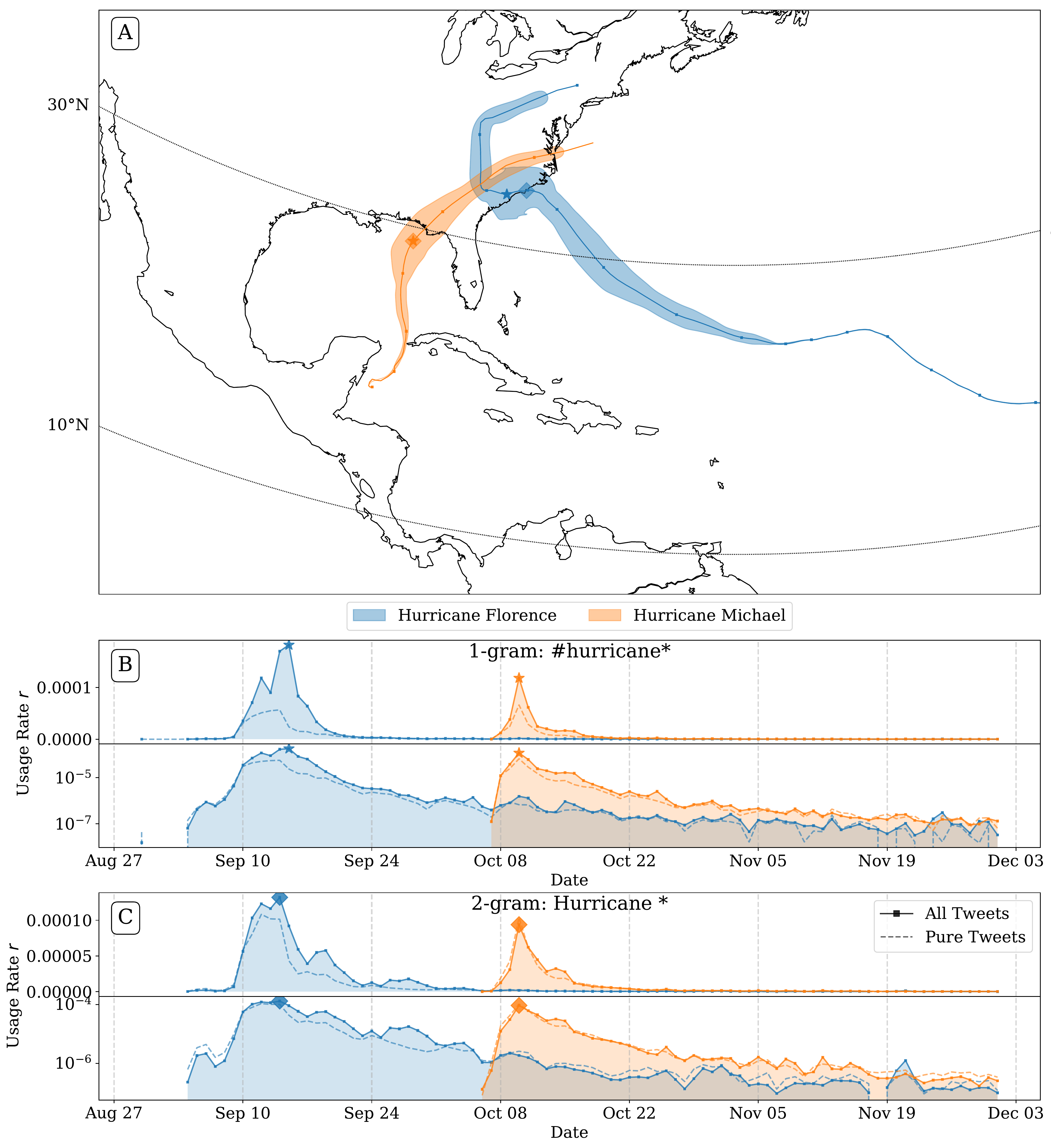}  
\caption{Hurricane Attention Map and time series Map and time series for 2018}\label{fig:2018}
\end{figure*}
%\subsection{$2$-gram usage rates}
%
%
%\begin{figure*}[t]
%  \centering	
%    \includegraphics[width=\linewidth]{figures/hurricane_2grams.pdf}  
% \caption{Hurricane $2$-gram usage rate from 2011 season to present. $2$-gram usage rates are bounded above by the $1$-gram usage rate \texttt{``Hurricane''}. During peak %attention the $1$-gram sharply rises from a usage rate of less than $1/100000$ to around $1/1000$ words in English language tweets. Time series are shown for usage %rates in both all tweets, as well as the 'pure' corpus, which filters out retweeted text. Usage rates in the all corpus typically have slightly higher peaks, and higher %variance than usage rates for the filtered text, reflecting that tweets referencing hurricanes are retweeted more than average during events.}
%  \label{fig:2gram_:q:rates}
%\end{figure*}

%subsection{Hashtag Ego Network}
%begin{figure*}[t]
% \centering	
%   \includegraphics[width=\linewidth]{figures/hurricaneharvey290.pdf}  
% \caption{
% \todo{Recommend removing this and saving for future work.}
%  The  ego network of \texttt{\#hurricaneharvey } for a single 12 hour period at the peak usage rate for for the central node on August 29, 2017 from  4am EST to 4pm EST. The network is a weighted co-occurrence network. Edge color represents the number of tweets in which two hashtags co-occurred. Node size represents the PageRank value for each node. Node color represents the usage rate of each hashtag. Here the hashtag with the largest usage rate is \texttt{\#Harvey}
% }
% \label{fig:ego_network}
%end{figure*}

\end{document}